\documentclass[%
superscriptaddress,
 amsmath,amssymb,
 twocolumn,
 aps,
prb,
]{revtex4-2}

\usepackage{graphicx}
\usepackage{dcolumn}
\usepackage{bm}
\usepackage{lineno}
\usepackage[colorinlistoftodos]{todonotes}
\usepackage{xcolor}
\usepackage{array}

\begin{document}

\title{Spin-forbidden excitations in the magneto-optical\\ spectra of CrI$_3$ tuned by covalency}

\author{Connor A. Occhialini}
\email{caocchia@mit.edu, Present address: Department of Physics, Columbia University, New York, NY 10027, USA}
\affiliation{Department of Physics, Massachusetts Institute of Technology, Cambridge, MA 02139, USA}

\author{Luca Nessi}
\affiliation{Department of Physics, Massachusetts Institute of Technology, Cambridge, MA 02139, USA}

\author{Luiz G. P. Martins}
\affiliation{Department of Physics, Massachusetts Institute of Technology, Cambridge, MA 02139, USA}
\affiliation{Department of Physics, Harvard University, Cambridge, MA 02138, USA}

\author{Ahmet Kemal Demir}
\affiliation{Department of Physics, Massachusetts Institute of Technology, Cambridge, MA 02139, USA}

\author{Qian Song}
\affiliation{Department of Physics, Massachusetts Institute of Technology, Cambridge, MA 02139, USA}

\author{Vicky Hasse}
\affiliation{Max Planck Institute for Chemical Physics of Solids, Dresden, Germany 01187}

\author{Chandra Shekhar}
\affiliation{Max Planck Institute for Chemical Physics of Solids, Dresden, Germany 01187}

\author{Claudia Felser}
\affiliation{Max Planck Institute for Chemical Physics of Solids, Dresden, Germany 01187}

\author{Kenji Watanabe}
\affiliation{Research Center for Functional Materials, National Institute for Materials Science, Tsukuba, Japan}

\author{Takashi Taniguchi}
\affiliation{International Center for Materials Nanoarchitectonics, National Institute for Materials Science, Tsukuba, Japan}

\author{Valentina Bisogni}
\affiliation{National Synchrotron Light Source II, Brookhaven National Laboratory, Upton, NY 11973, USA}

\author{Jonathan Pelliciari}
\affiliation{National Synchrotron Light Source II, Brookhaven National Laboratory, Upton, NY 11973, USA}

\author{Riccardo Comin}
\email{rcomin@mit.edu}
\affiliation{Department of Physics, Massachusetts Institute of Technology, Cambridge, MA 02139, USA}

\date{\today}

\begin{abstract} 
Spin-forbidden ($\Delta S \neq 0$) multiplet excitations and their coupling to magnetic properties are of increasing importance for magneto-optical studies of correlated materials. Nonetheless, the mechanisms for optically brightening these transitions and their generality remain poorly understood. Here, we report magnetic circular dichroism (MCD) spectroscopy on the van der Waals (vdW) ferromagnet (FM) CrI$_3$. Previously unreported spin-forbidden ($\Delta S = 1$) ${}^4A_{2\mathrm{g}} \to{}^2E_\mathrm{g}/{}^2T_{1\mathrm{g}}$ Cr${}^{3+}$ $dd$ excitations are observed near the ligand-to-metal charge transfer (LMCT) excitation threshold. The assignment of these excitations and their Cr$^{3+}$ multiplet character is established through complementary Cr $L_3$-edge resonant inelastic X-ray scattering (RIXS) measurements along with charge transfer multiplet (CTM) calculations and chemical trends in the chromium trihalide series (CrX$_3$, X = Cl, Br, I). We utilize the high sensitivity of MCD spectroscopy to study the thickness dependent optical response. The spin-forbidden excitations remain robust down to the monolayer limit and exhibit a significant magnetic field tunability across the antiferromagnetic to FM transition in few-layer samples. This behavior is associated to changes in the metal-ligand covalency with magnetic state, as supported by our CTM analysis. Our results clarify the magneto-optical response of CrI$_3$ and identify covalency as a central mechanism for the brightening and field-tunability of spin-forbidden multiplet excitations.

\end{abstract}

\maketitle

\section{Introduction}

Chromium trihalides (CrX$_3$, X = Cl, Br, I) are van der Waals (vdW) layered magnetic insulators \cite{Hansen1959}. CrBr$_3$ and CrI$_3$ are historically important as they are among the first known examples of insulating ferromagnets \cite{Hansen1959, Tsubokawa1960, Dillon1962, Dillon1963, Dillon1966, Krinchik1969, Day1978}. Thus, there has been significant effort in understanding the magneto-optical spectra of these materials \cite{Dillon1962, Dillon1963, Dillon1965, Dillon1966, Pedroli1975, Carricaburu1986, Pollini1989, Bermudez1979i, Bermudez1979ii, Grant1968}. In the more ionic cases of CrBr$_3$ and CrCl$_3$, the optical range excitations ($E \simeq 1.5 \to 3$ eV) are well described by Cr${}^{3+}$ ($3d^3$) $dd$ excitations \cite{Bermudez1979i, Bermudez1979ii, Dillon1966, Pedroli1975} within ligand field theory (LFT) \cite{ballhausen1962}, followed by ligand-to-metal charge transfer (LMCT) transitions at higher energy ($>$ 2.6 and $>$ 3.6 eV for X $=$ Br/Cl, respectively) \cite{Pedroli1975, Carricaburu1986, Pollini1989, Shinagawa1996}. Meanwhile, the reduced charge transfer energy and increased metal-ligand covalency for CrI$_3$ challenges such assignments, with both LMCT and $dd$ transitions overlapping in the near infrared/visible energy range ($1.5 \to 3.0$ eV) \cite{Grant1968}. The optical spectra and electronic energy scales of CrI$_3$ have received relatively less attention until recently \cite{McGuire2015,Huang2017b,Seyler2018,Song2021,  DeSiena2020, Jin2020, Acharya2022, Galbiati2023, Frisk2018, Choi2020, Ghosh2023, Shao2021}.

CrX$_3$ -- in particular CrI$_3$ -- are of strong contemporary interest in the context of two-dimensional (2D) magnetism.
CrI$_3$ is an easy-axis FM ($T_\mathrm{C} \sim 61$ K \cite{McGuire2015}) that is exfoliable and maintains long-range intralayer FM order down to the single-layer limit \cite{Huang2017b,Zhang2022}. This has enabled the study of emergent magnetic properties in 2D, including the layer-dependent magnetic order where few-layer samples become antiferromagnetic (AFM) due to structurally-induced changes in the interlayer magnetic exchange \cite{Huang2017b,Guo2021,Li2020,Jang2019,McGuire2015}. The highly tunable interlayer exchange has been key for studies of novel magnetic states in, e.g., moir\'{e} heterostructures \cite{Song2021b, Xu2022, Xie2023}.

Central to many of these studies are magneto-optical effects (e.g., magnetic circular dichroism [MCD] and the magneto-optic Kerr effect [MOKE]), which provide the primary methods to probe the magnetic order of atomically thin samples. Modern works favor the interpretation of magneto-optical effects within band structure plus Bethe-Salpeter equation (BSE) approaches \cite{Wu2019,Molina2020,Wu2022,Acharya2022}. However, the connection between BSE and the ligand field/molecular orbital theory that appropriately describes CrBr$_3$/CrCl$_3$ is not transparent, leading to inconsistencies in the interpretation. Meanwhile, the magneto-optical spectra have not been systematically investigated in CrI$_3$ and it is unknown how they depend on layer number and the underlying magnetic state. Thus, despite the importance for probing magnetic order \cite{Huang2017b} and the mechanisms of photoinduced magnetic effects \cite{Padmanabhan2022, Dabrowski2022, Zhang2022, Grzeszczyk2023}, an understanding of the magneto-optical response of CrI$_3$ is currently incomplete.

There has also been recent interest in spin-forbidden excitations contributing to the magneto-optical response. This is exemplified by the spin-flip ${}^2E_\mathrm{g}$/${}^2T_{1\mathrm{g}} \to {}^4A_{2\mathrm{g}}$ luminescence in CrPS$_4$ \cite{Gu2020,Kim2022,Multian2024} -- commonly observed in Cr${}^{3+}$ compounds \cite{Sugano1958,Liehr1963,Macfarlane1963,Dillon1966,Pedroli1975,Bermudez1979ii,Schmidt2013,Rabia2014, Macfarlane1971} -- and the sharp ${}^1A_{1\mathrm{g}} \leftrightarrow {}^3A_{2\mathrm{g}}$ excitations in optical absorption and luminescence of NiPS$_3$ \cite{Banda1986, Kang2020,Hwangbo2021, Wang2021, Ergecen2022, Jana2023, Kim2023, He2024} and the nickel dihalides \cite{Banda1986, Kozielski1972,Giordano1978,Robbins1976,Rosseinsky1978,Son2022, Occhialini2024, Lebedev2024}. Spin-forbidden excitations correspond to a change the total spin ($S$) of the transition-metal (TM) ion and are associated to a specific type of $dd$ excitation with energies determined by Hund's coupling \cite{Sugano1970,Lohr1972,Kitzmann2022}. Such excitations are of interest due to their characteristically small linewidths \cite{Lohr1972} and the proposed connection of their optical cross section to magnetic order \cite{Robbins1976,Kozielski1972,Giordano1978,Eremenko1977,Kang2020} or inversion symmetry breaking \cite{Ergecen2022,Son2022,Multian2024}. Thus, such excitations may provide a means to optically probe the magnetic properties of 2D materials. However, the generality, properties, and mechanisms for enabling spin forbidden transitions have not been clarified. In particular, such excitations have not been identified in CrI$_3$, in part due to the complexity of the optical spectra.

In this work, we report MCD spectroscopy and Cr $L_3$-edge resonant inelastic X-ray scattering (RIXS) studies of CrI$_3$. MCD spectroscopy measurements reveal the presence of both spin-allowed (${}^4A_{2\mathrm{g}} \to {}^4T_{2\mathrm{g}}/{}^4T_{1\mathrm{g}}$) and unreported spin-forbidden (${}^4A_{2\mathrm{g}} \to {}^2E_\mathrm{g}/{}^2T_{1\mathrm{g}}$) excitations near the lowest allowed LMCT band edge. The excitations are assigned through RIXS measurements, charge transfer multiplet (CTM) calculations, and chemical trends in the CrX$_3$ series. We further use the high sensitivity of MCD spectroscopy to study the thickness dependence of the optical spectra. These data demonstrate that the spin-forbidden excitations remain robust down to the monolayer limit, with a significant magnetic field tunability observed in antiferromagnetic few-layer samples. Through comparison with the MCD spectrum of CrBr$_3$, we associate these observations to an increased Cr-X covalency that is tunable by both ligand and the magnetic state. These results yield a consistent interpretation of the magneto-optical spectra of the chromium trihalides and underscore the importance of metal-ligand covalency for the mechanisms of brightening and tuning spin-forbidden multiplets in correlated insulators.

\section{Results}

\begin{figure}
\centering

\includegraphics[width = 1.00\columnwidth]{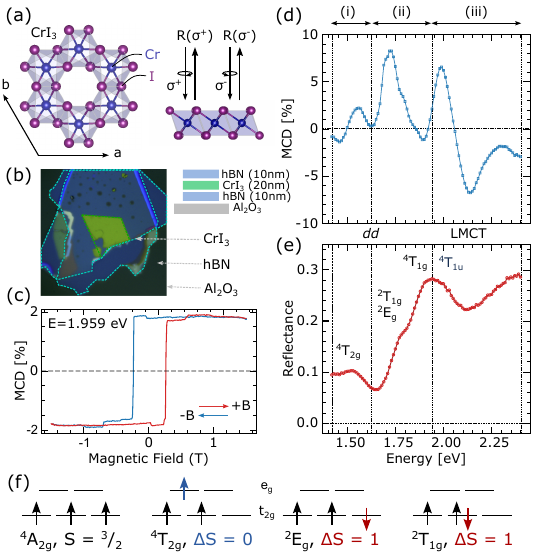}

\caption{(a) Planar crystal structure of CrI$_3$ and MCD measurement geometry with $\sigma^+$/$\sigma^-$ polarizations incident along the $\hat{c}$ axis. (b) $\sim 20$ nm thick CrI$_3$ flake (green outline) double encapsulated by $\sim 10$ nm hBN (blue) transferred onto an Al$_2$O$_3$ substrate for optical measurements. (c) MCD hysteresis loop at $E = 1.959$ eV and $T = 1.75$ K. (d) MCD and (e) reflectance ($R(\epsilon)$) spectra at $T = 1.75$ K. The different spectral regions (i)-(iii) and peak assignments are indicated (see text for details). (f) Schematics of the ground and lowest excited state multiplets for the Cr$^{3+}$ ($d^3$) ion in octahedral symmetry. The crystal structure in (a) is generated by VESTA \cite{Momma2008}.}

\label{fig:fig1}
\end{figure}

\subsection{Magneto-optical spectra of bulk CrI$_3$}

We begin by discussing the MCD  [MCD($\epsilon$)] and reflectance [$R(\epsilon)$] spectra versus photon energy ($\epsilon$) in bulk CrI$_3$ recorded in back scattering geometry at normal incidence [Fig. \ref{fig:fig1}(a)] (for experimental details, see Appendix). Bulk spectra were recorded on a $\sim 20$ nm thick flake double encapsulated with hBN [Fig. \ref{fig:fig1}(b)]. At $T = 1.75$ K, we first recorded an MCD hysteresis loop at $E = 1.959$ eV and $\vec{B} \parallel \hat{c}$ along the easy axis of CrI$_3$, confirming FM behavior with coercive field $B_\mathrm{c} \sim 250$ mT [Fig. \ref{fig:fig1}(c)]. MCD($\epsilon$) and $R(\epsilon)$ are reported in Figures \ref{fig:fig1}(d)/(e), respectively. The MCD spectra are recorded at $B = 0$ T and antisymmetrized with respect to the magnetic state by applying $\pm B_\mathrm{c}$ and returning to $B = 0$ T.

We observe several prominent features in MCD($\epsilon$) and $R(\epsilon)$ measurements including transitions at the following energies: (i) $E =$ 1.4 $\to$ 1.62 eV, (ii) 1.62 $\to$ 1.94 eV and (iii) 1.94 $\to$ 2.4 eV [Fig. \ref{fig:fig1}(d)/(e)]. The lowest energy feature in MCD($\epsilon$) and $R(\epsilon)$ agrees with the lowest energy Cr$^{3+}$ multiplet term in $O_\mathrm{h}$ CF, representing the (spin-allowed, $\Delta S = 0$) ${}^4A_{2\mathrm{g}} \to {}^4T_{2\mathrm{g}}$ excitation, associated to a $t_{2\mathrm{g}} \to e_\mathrm{g}$ inter-orbital transition \cite{Seyler2018} [Fig. \ref{fig:fig1}(f)]. At higher energies, a larger reflectance band edge is observed beginning around 1.7 eV and peaking near 1.94 eV. The peak near 1.94 eV is a LMCT transition \cite{Grant1968,Shinagawa1996,Seyler2018}, mixed with the higher energy spin-allowed ${}^4T_{1\mathrm{g}}$ multiplet as shown below. 

\begin{figure}
\centering

\includegraphics[width = \columnwidth]{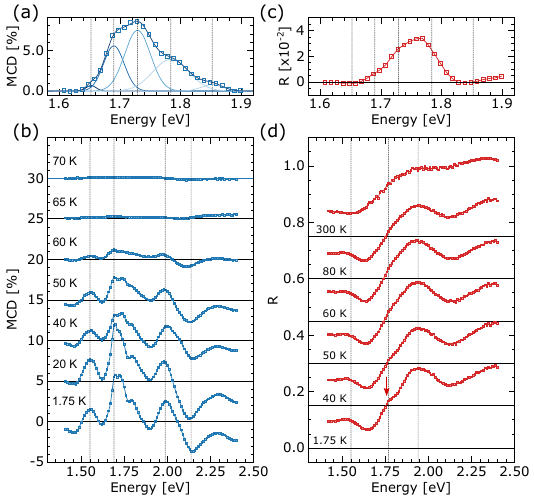}

\caption{(a) Zoom in of the MCD spectrum in the energy range 1.6 - 1.9 eV at $T$ = 1.75 K on the 20 nm sample in Fig. \ref{fig:fig1}. Black squares are data and the black line is a fit with a linear baseline subtracted and Gaussian fit components shown in blue. (b) Temperature-dependence of the MCD spectrum showing the disappearance across $T_\mathrm{C} \simeq 60$ K, as expected. (c) Zoom in of the shoulder of $R(\epsilon)$ at $T$ = 1.75 K in the energy range 1.6 - 1.9 eV with nearby spectral components subtracted from a minimal Gaussian peak fit. (d) Temperature-dependence of $R(\epsilon)$. Vertical dashed lines in (a)-(d) are guides to the eye. Red arrow in (d) highlights the $T$-dependent fine structure component near 1.8 eV. }

\label{fig:fig2}
\end{figure}

Preceding the LMCT feature, we observe a prominent multi-peaked excitation from $1.65 \to 1.90$ eV in MCD($\epsilon$). Previous optical measurements have observed excitations in this energy range, but without assignment \cite{Seyler2018,Song2021,DeSiena2020,Jin2020}. A zoom-in of this spectral region is shown in Fig. \ref{fig:fig2}(a), showing multiple peaks between 1.650-1.850 eV which appear qualitatively sharper than other features in the MCD spectrum. A corresponding reflectance signal is also observed in a similar energy range [Fig. \ref{fig:fig2}(c)/(d)]. The lowest energy peaks in MCD($\epsilon$) agree well with the expected energy of ${}^2E_\mathrm{g}/{}^2T_{1\mathrm{g}}$ spin-forbidden ($\Delta S = 1$) transitions [Fig. \ref{fig:fig1}(f)], which are commonly observed in the luminescence of strong field octahedrally-coordinated ($O_\mathrm{h}$) Cr$^{3+}$ ions/color centers \cite{Sugano1958,Macfarlane1971,Liehr1963}, while the higher energy peaks resemble two-phonon/magnon sidebands \cite{Kozielski1972, Robbins1976, Macfarlane1971, Dillon1966, Pedroli1975, Bermudez1979ii}. The precise assignment of all features in this spectral region (in the bulk) is complicated by the presence of both resonance and anti-resonance lineshapes in MCD spectra and will be substantiated by additional measurements below. However, these results reveal sharp excitations near 1.7 eV and show that MCD spectroscopy is effective to resolve them from the more strongly allowed LMCT bands.

We proceed with temperature dependence of MCD($\epsilon$) and $R(\epsilon)$ as reported in Fig. \ref{fig:fig2}(b,d). The zero-field MCD signal continuously reduces to zero with increasing $T$ up until $T_\mathrm{C} \simeq 60$ K, as expected with the vanishing ordered magnetization \cite{Dillon1965,Hansen1959}. In the $R(\epsilon)$ measurements [Fig. \ref{fig:fig2}(d)], a  significant temperature dependence is observed for the 1.8 eV shoulder near the LMCT transition, which broadens and eventually merges with nearby excitations with increasing temperature. This fine structure has been associated with either a bright exciton from BSE calculations or the ${}^4T_{1\mathrm{g}}$ (spin-allowed) $dd$ excitation \cite{Grant1968,Seyler2018,Jin2020,Wu2019}. We examine this energy region more closely by fitting the ${}^4T_{2\mathrm{g}}$ and LMCT transitions with Gaussian peaks and subtracting them from the signal to report the additional 1.8 eV spectral weight in Fig. \ref{fig:fig2}(c). A close correspondence between this excitation with the side band region of the $\Delta S = 1$ transitions observed in MCD [Fig. \ref{fig:fig2}(a)] is noted, and we assign this feature accordingly. This assignment may explain the significant vibronic coupling observed in Raman measurements with incident energy tuned to this spectral region \cite{Jin2020}.

\begin{figure*}
\centering

\includegraphics[width = \textwidth]{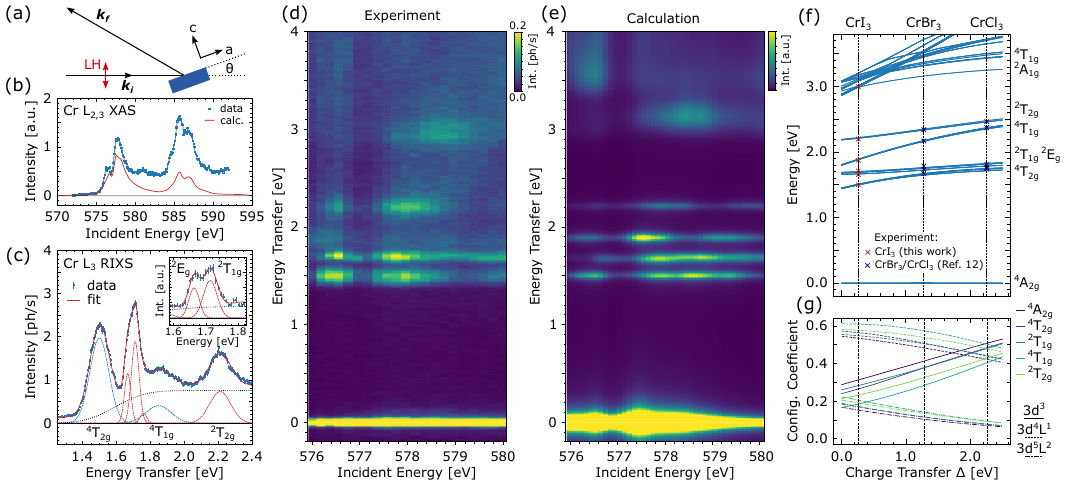}

\caption{(a) Experimental geometry for RIXS measurements. (b) Cr $L_{2,3}$-edge XAS measurements in PFY (blue) at $\theta = 90^\circ$ with the calculated XAS from CTM modeling overlaid (red, see text). (c) Cr $L_3$-edge RIXS spectrum integrated over incident energy from the (d) energy-dependent RIXS map. In (c), blue squares are the experimental data with individual Gaussian fit components to $\Delta S = 0$ (blue) and $\Delta S = 1$ (red) excitations shown. The black line is a step-like background to account for fluorescence. Inset: a zoom in to the region 1.6-1.8 eV recorded in a higher resolution ($\Delta E \simeq 22$ meV) at $E = 577.5$ eV, highlighting the ${}^2E_\mathrm{g}/{}^2T_{1\mathrm{g}}$ excitations. Comparison between (d) experimental and (e) calculated RIXS maps recorded at $T = 40$ K, $\theta = 15^\circ$ and $2 \theta = 150^\circ$ with incident $\pi$ polarization. The calculation considers the same experimental geometry and polarization conditions. (f) Energy level diagram as a function of the charge transfer gap ($\Delta$) based on optimized fits to the lowest energy multiplet excitations in CrI$_3$, CrBr$_3$ and CrCl$_3$. The energies for CrBr$_3$ and CrCl$_3$ are from optical absorption in Ref. \cite{Bermudez1979i} (purple crosses) and CrI$_3$ values are determined from RIXS measurements in panel (c) (red crosses). (g) Evolution of the $3d^3$, $3d^4\underline{L}^1$ and $3d^5\underline{L}^2$ configurational weights resolved to the ground and lowest excited state multiplets as a function of $\Delta$. In (f) and (g), the ionic crystal field splitting ($10Dq$) and the nephelauxetic reduction ($\beta$) are covaried with $\Delta$ based on interpolation of the optimized values as a function of the halogen (see text and Table \ref{tab:tab1}).}

\label{fig:fig3}
\end{figure*}

\subsection{Cr $L_3$-edge RIXS measurements}

In order to bolster the assignments of the excitations in the optical spectra, we performed RIXS measurements at the Cr $L_3$-edge ($2p_{3/2} \to 3d$) on bulk CVT-grown single crystals of CrI$_3$. The distinct selection rules of the two-dipole transition RIXS process alongside the spin-orbit coupling (SOC) in the core levels allow direct sensitivity to both spin-allowed and spin-forbidden $dd$ excitations of transition metal ions \cite{DeGroot2008}. Furthermore, local cluster models based on ligand field or charge transfer multiplet theory are able to quantitatively capture the excitation energies and their intensity/energy-dependence \cite{Cowan1981, Haverkort2012}, providing additional information for peak assignment. 

RIXS measurements were performed at the 2-ID (SIX) beamline at the National Synchrotron Light Source II (NSLS-II) \cite{Dvorak2016}, with a resolution of $\Delta E \sim 27$ meV and at a temperature of $T = 40$ K. We utilized a fixed scattering geometry with $2\theta = 150^\circ$ scattering angle and incident linear horizontal (LH, $\pi$) polarization [Fig. \ref{fig:fig3}(a)]. The samples were handled and cleaved inside an N$_2$ environment before loading into the high vacuum sample chamber.

Figure \ref{fig:fig3}(b) shows the $L_{2,3}$-edge XAS spectrum of CrI$_3$, recorded in partial fluorescence yield (PFY) with $\pi$ incident polarization at normal incidence ($\theta = 90^\circ$). We measured incident energy dependent RIXS maps across the Cr $L_3$-edge at fixed $\theta = 15^\circ$ with incident $\pi$  polarization [Fig. \ref{fig:fig3}(d)]. We observe a sequence of sharp Raman-like excitations in the energy range $1.5 \to 3.0$ eV. The excitations in the $1.3 \to 2.4$ eV energy range can be observed more clearly in Fig. \ref{fig:fig3}(c) where we report the incident-energy-integrated RIXS spectrum (from 576.0 to 580.0 eV). We fit the sharp features and identify excitations at energies of 1.497 eV (131 meV), 1.665 eV (45 meV), 1.709 eV (53 meV), 1.753 eV (45 meV), 1.846 eV (175 meV), and 2.211 eV (147 meV) with the values in parentheses representing the FWHM from Gaussian fits. The doublet peak near 1.7 eV is qualitatively sharper than the other excitations similar to observations in MCD [Fig. \ref{fig:fig2}(a)]. A higher resolution measurement with $\Delta E = 22$ meV at incident energy $E = 577.5$ eV is shown in Fig. \ref{fig:fig3}(c) (inset), showing a well-resolved doublet excitation with energies (FWHM) of 1.664 eV (38 meV) and 1.712 eV (48 meV). Each feature below 2.0 eV in the RIXS spectrum has a corresponding excitation below the LMCT threshold in MCD($\epsilon$) and $R(\epsilon)$ measurements [Figs. \ref{fig:fig1}/\ref{fig:fig2}].

\subsection{CrX$_6$ cluster calculations}

To assign the features in the RIXS and optical experiments, we utilize charge transfer multiplet (CTM) theory and model the Cr $L_{2,3}$-edge XAS/RIXS spectra using the QUANTY software \cite{Haverkort2012, Haverkort2014}. The large Cr-I covalency \cite{Occhialini2024} necessitates the  consideration of charge transfer effects including the $3d^3$, $3d^4\underline{L}$ and $3d^5\underline{L}^2$ electronic configurations in the calculations, where $\underline{L}$ corresponds to a hole in the ligand orbitals (I-$5p$). We utilize the standard procedure \cite{Haverkort2012} and consider a single CrI$_6$ cluster with basis consisting of the five Cr-$3d$ orbitals and the symmetrized (bonding) I-$5p$ molecular orbitals of $e_\mathrm{g}$ and $t_{2\mathrm{g}}$ symmetry. We consider the case of $O_\mathrm{h}$ symmetry. 

The parameters in the cluster model include the Cr $3d$-$3d$ and $2p$-$3d$ Coulomb interactions (Slater integrals), the Cr $2p$/$3d$ spin orbit coupling (SOC), the $O_\mathrm{h}$ crystal field splitting ($10Dq$), the hybridization between the metal and ligand $e_\mathrm{g}$ and $t_{2\mathrm{g}}$ states, the charge transfer gap ($\Delta$), and the Coulomb repulsion in the initial and intermediate RIXS states ($U_{dd}$/$U_{pd}$, resp.). The key parameters for the present discussion are the crystal field splitting ($10Dq$), the charge transfer gap ($\Delta$) and the nephelauxetic reduction ($\beta$) which represents the reduction of Slater integrals from atomic values. Further details of the parameters are provided in the Appendix. 

The optimized results of the calculations for CrI$_3$ are shown in Fig. \ref{fig:fig3}(b) for the XAS spectrum (red line) with the corresponding RIXS map shown in Fig. \ref{fig:fig3}(e). The parameters are reported in Table \ref{tab:tab1}. We find an excellent agreement for both data sets, including the energy of all observed Raman-like excitations in the RIXS profile and their incident energy dependence. The relative intensity deviations between the Cr $L_3$ pre- and main-edge in RIXS, as well as the $L_2$/$L_3$ branching ratio in XAS, are attributable to self-absorption effects \cite{DeGroot2008}. This allows the $dd$ excitation features to be assigned as: ${}^4T_{2\mathrm{g}}$ (1.5 eV), ${}^2E_\mathrm{g}$ (1.664 eV), ${}^2T_{1\mathrm{g}}$ (1.712 eV), ${}^4T_{1\mathrm{g}}$ (1.85 eV), ${}^2T_{2\mathrm{g}}$ (2.21 eV) and ${}^4T_{2\mathrm{g}}$ (3.0 eV). Most importantly, the doublet excitation revealed in MCD measurements near 1.7 eV [Fig. \ref{fig:fig1}/\ref{fig:fig2}] corresponds to the spin flip $\Delta S = 1$ intra-configurational ${}^2E_\mathrm{g}$/${}^2T_{1\mathrm{g}}$ multiplets.

To substantiate the accuracy of the CTM model, we take the observed $dd$ excitation energies for CrBr$_3$ and CrCl$_3$ below $3$ eV from the available optical literature \cite{Bermudez1979i} and fit the excitations to the same model. The optimized parameters for each ligand are reported in Table \ref{tab:tab1} with the corresponding energy level diagram as a function of $\Delta$ shown in Fig. \ref{fig:fig3}(f). A satisfactory fit for all ligands is achieved for the multiplets below 3 eV. The fitted values of $\Delta$ are systematically lower than results from X-ray photoelectron spectroscopy (XPS) analysis \cite{Pollini1999}, since $\Delta$ reported here refers to the center of the configurations and not with respect to the ${}^4A_{2\mathrm{g}}$ ground state after including multiplet coupling and the crystal field \cite{Zaanen1985}. However, the trend of $\Delta$ we optimize across the series is in good agreement with reports from both XPS \cite{Pollini1999} and optical spectroscopy of the charge transfer band edge \cite{Pollini1989, Shinagawa1996}.

\begin{table}[th!]
\centering

\begin{tabular}{ |c || wc{1.75cm}|wc{1.75cm}|wc{1.75cm}||} 
 X & $\beta$ & $10Dq$ [eV] & $\Delta$ [eV]\\ [0.5ex] 
 \hline\hline
 I  & 0.906 & 1.300 & 0.26 \\ 
 Br & 0.916 & 1.445 & 1.28 \\
 Cl & 0.925 & 1.500 & 2.27 \\
\end{tabular}

\caption{Optimal cluster model parameters for CrX$_3$ for X = I, Br, Cl considering the $3d^3$/$3d^4\underline{L}$/$3d^5\underline{L}^2$ configurations (see text for details). 
}

\label{tab:tab1}
\end{table}

To examine the role of charge transfer, we extract the configurational contributions to the ground and lowest energy excited states versus $\Delta$ in Fig. \ref{fig:fig3}(g). This shows an excitation-dependent mixing between the ionic and charge transfer  configurations, with significant $3d^5\underline{L}^2$ configurational weight observed -- particularly for CrI$_3$ ($\Delta = 0.26$ eV). Restricting the model to only the $3d^3$ and $3d^4\underline{L}$ configurations resulted in inaccurate fits to the extreme case of CrI$_3$ and required significantly smaller values of $\Delta$ for all ligands. Similarly, the atomic ($3d^3$) results (e.g., Tanabe Sugano diagrams \cite{ballhausen1962, Shao2021}) are not able to reproduce the experimental energy levels. Thus, configuration interactions with charge transfer states are crucial to describe the observed excitations. Some features have an overestimated intensity in the CTM calculation, in particular the ${}^4T_{1\mathrm{g}}$ excitation [Fig. \ref{fig:fig3}(d,e)]. This overestimation correlates with the excitation-dependent contribution of the charge transfer configurations [Fig. \ref{fig:fig3}(g)]. We associate this to the effects of the ligand bandwidth which is not included in the calculations. This suggests that single cluster models are at the margins of applicability in the case of CrI$_3$ with band structure effects becoming important \cite{Acharya2022}. 

We finally address the relationship between the charge transfer configurations and the LMCT excitons observed in optical experiments (Fig. \ref{fig:fig1}) \cite{Pollini1989}. The lowest energy dipole-allowed transitions in optics are from the non-bonding $t_{1\mathrm{u}}$/$t_{2\mathrm{u}}$ molecular orbitals to the (anti-bonding, metal-centered) $e_\mathrm{g}^*$ states with term ${}^4T_{1\mathrm{u}}$ \cite{Shinagawa1996}. Being non-bonding, the $t_{1\mathrm{u}}$ and $t_{2\mathrm{u}}$ ligand orbitals are (by construction) absent from the cluster model which only considers the bonding $e_\mathrm{g}$ and $t_{2\mathrm{g}}$ ligand orbitals \cite{Haverkort2012}. Therefore, the LMCT excitons observed at 1.94 eV and 2.8 eV in optics  \cite{Grant1968,Seyler2018,Shinagawa1996} are not included in the model and are distinct from the $dd$ transitions observed and calculated at 2.21 and 3.0 eV. The LMCT excitons are more strongly dipole-allowed and likely obscure the observation of these higher energy $dd$ features in optics. The ${}^4T_{1\mathrm{u}}$ excitations are not clearly observed in RIXS, but the multiplet assignments show a near degeneracy between the experimentally observed LMCT exciton and the ${}^4T_{1\mathrm{g}}$ $dd$ excitation, confirming the assignments in Fig. \ref{fig:fig1}.

\begin{figure}
\centering

\includegraphics[width = \columnwidth]{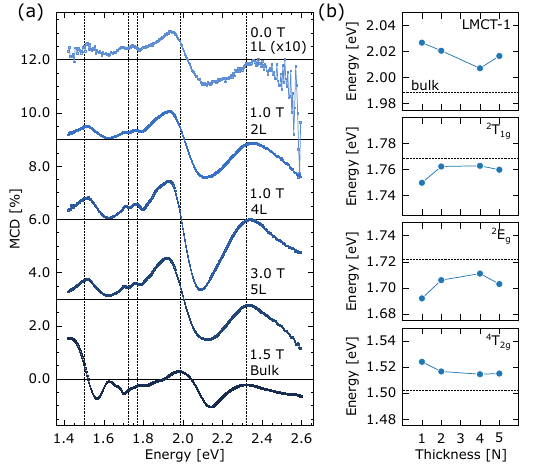}

\caption{(a) Thickness dependent MCD spectra at $T = 1.75$ K for 1L, 2L, 4L, 5L and exfoliated bulk (thickness $\sim 100$ nm). Spectra are recorded at the indicated field to achieve the fully saturated FM state in each sample. Vertical lines are guides to the eye. (b) Thickness dependence of the energies for the principle $dd$ and LMCT transitions. Horizontal dashed lines are the bulk values from optical measurements.}

\label{fig:fig4}
\end{figure}

\subsection{Layer-dependent magneto-optical spectra}

Having established the origin of the optical excitations of bulk CrI$_3$, we utilize the high sensitivity of MCD spectroscopy to the spin forbidden ${}^4A_{2\mathrm{g}} \to {}^2E_\mathrm{g}/{}^2T_{1\mathrm{g}}$ excitations to study the their evolution with sample thickness. Layer dependent magneto-optical spectra are reported in Fig. \ref{fig:fig4}(a), including 1L, 2L, 4L, 5L and an additional exfoliated bulk sample ($\sim$ 100 nm thickness). All spectra are recorded from samples on Al$_2$O$_3$ substrates. Each spectrum is taken at sufficiently large magnetic field to saturate the magnetization, which depends on the layer-number due to the antiferromagnetic (AFM) interlayer exchange coupling in few-layer CrI$_3$ \cite{Huang2017b}. The few-layer data are qualitatively distinct from the bulk due to finite thickness (thin film) effects in the latter leading to anti-resonance lineshapes for, e.g., the ${}^4T_{2\mathrm{g}}$ excitation in MCD [Fig. \ref{fig:fig1}(d), \ref{fig:fig4}(a)]. The few-layer data instead closely resemble previously reported bulk-like MCD spectra in transmission \cite{DeSiena2020}. Despite details of the spectral shape, the key excitations in the bulk (Fig. \ref{fig:fig1}) are also observed in all thicknesses down to the monolayer, including the $\Delta S = 1$ excitations near 1.7 eV. 

We extract the layer dependent excitation energies in Fig. \ref{fig:fig4}(b).  A small blueshift (redshift) of the ${}^4T_{2\mathrm{g}}$/${}^4T_{1\mathrm{u}}$ (${}^2E_\mathrm{g}$/${}^2T_{1\mathrm{g}}$) excitations of $<30$ meV from bulk to monolayer is observed, respectively. This could be due to small renormalization of $\Delta$, $10Dq$ and/or the electronic correlations with thickness as previously argued in other 2D magnets like NiPS$_3$ \cite{Wang2022,DiScala2024}, and agrees for the LMCT transition shift observed at room temperature \cite{Galbiati2023}, though with a smaller magnitude in our data. Crucially, the magneto-optical spectra and the spin-forbidden multiplets are robust down to the monolayer limit.

\begin{figure}
\centering

\includegraphics[width = \columnwidth]{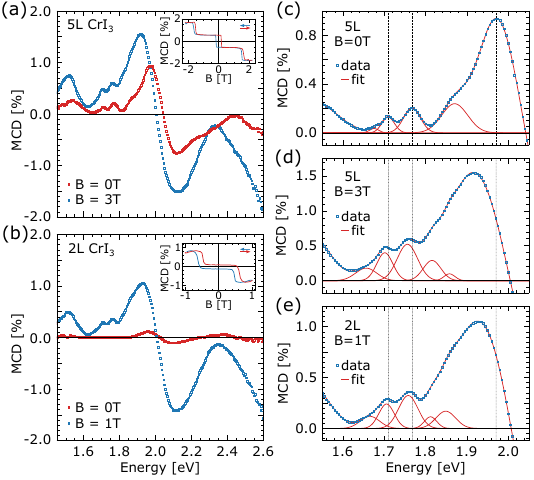}

\caption{Dependence of the MCD spectrum on magnetic state in few layer CrI$_3$ at $T = 1.75$ K. (a) Antisymmetrized MCD spectrum for 5L CrI$_3$ at $B = 0.0$ T (red) and $B = 3.0$ T (blue), respectively. (b) Antisymmetrized MCD spectrum for 2L CrI$_3$ at $B = 0.0$ T (red) and $B = 1.0$ T (blue), respectively. Insets to (a)/(b) are the rsepective hysteresis loops recorded with fixed $E = 1.959$ eV. Fits to the 1.6-2.0 eV region highlighting the $\Delta S = 1$ excitations for (c) 5L CrI$_3$ at $B = 0.0$ T, (d) 5L at $B = 3.0$ T and (e) 2L at $B = 1.0$ T. Blue squares are the experimental data and red solid lines indicate the total and individual fit components.}

\label{fig:fig5}
\end{figure}

As noted in the bulk [Fig. \ref{fig:fig2}], the ${}^2E_\mathrm{g}$/ ${}^2T_{1\mathrm{g}}$ peaks in MCD are systematically higher in energy compared to RIXS [Fig. \ref{fig:fig3}(c)], with an energy difference of around 40-50 meV. This could be due to a difference in MCD and absorption, where the position of maximum dichroism does not necessarily align to the fundamental energy of the underlying excitation. Alternatively, these peaks could be two-phonon/magnon side bands \cite{Macfarlane1971, Schmidt2013, Benedek1979}, consistent with the magnon/phonon dispersion bandwidth of $\sim 20$/$30$ meV \cite{Chen2018,Djurdjic2018}, respectively. In particular, two-magnon side bands are common to the spin forbidden excitations of Cr$^{3+}$ systems \cite{Schmidt2013}. This may partially explain their large contribution in MCD and their temperature dependent broadening above $T_\mathrm{C}$ [Fig. \ref{fig:fig2}(d)]. Close inspection (Fig. \ref{fig:fig5}) shows additional spectral weight near 1.65 eV. This may represent the fundamental (zero-boson) ${}^2E_\mathrm{g}$ excitation, as this aligns more closely with the ${}^2E_\mathrm{g}$ peak in RIXS [Fig. \ref{fig:fig3}(c)], though higher resolution optical studies would be required to ascertain whether any sharp zero-boson $\Delta S = 1$ excitations are present. However, we highlight that the linewidths in optics ($\sim 27$ meV) and in RIXS ($\sim 38/48$ meV) are both not resolution limited. This suggests that even the fundamental multiplet excitations are intrinsically broad.

\subsection{Magnetic-field dependence}

The layer-dependent ground state of few-layer CrI$_3$ offers additional opportunities to study the effects of the magnetic state on the optical spectrum that are inaccessible in bulk. While the bulk is FM, the weak interlayer AFM coupling of few-layer CrI$_3$ allows a switching between layered AFM/FM states through metamagnetic transitions \cite{Huang2017b, Jang2019,Li2020,Guo2021}. We thus studied the magnetic field dependent MCD spectrum of 5L and 2L CrI$_3$ at $T = 1.75$ K shown in Fig. \ref{fig:fig5}(a)/(b), respectively. The insets show the hysteresis at fixed $E = 1.959$ eV for each sample, confirming the field induced transition from AFM to FM stacking configurations at $B_\mathrm{c}(5L) \sim 1.8$ T and $B_c(2L) \sim 0.6$ T. The 5L sample shows a remnant magnetization at zero field due to the odd layer number while the bilayer is nearly compensated at $B = 0$ T, consistent with previous reports \cite{Song2018}.

Fig. \ref{fig:fig5}(a) compares the MCD spectrum at $B = 0$ T and $B = 3$ T in the 5L sample. Besides an increased overall MCD magnitude at higher field due to the increased net magnetization, qualitative differences are observed in the MCD spectrum in the AFM and FM coupled states. This includes a shift of the LMCT transitions to lower energy, along with a weaker redshift of the ${}^2E_\mathrm{g}$/${}^2T_{1\mathrm{g}}$/${}^4T_{2\mathrm{g}}$ multiplets. For the bilayer [Fig. \ref{fig:fig2}(b)], the zero field MCD contrast is small across the full spectral range, while a similar field-induced redshift of the LMCT transition is observed. We focus on the region of the LMCT transition and the $\Delta S = 1$ excitations in Fig. \ref{fig:fig5}(c)-(e) and fit the MCD spectrum. These quantify a field-induced redshift of $\sim 8$/$\sim 10$ meV for the ${}^2E_\mathrm{g}$/${}^2T_{1\mathrm{g}}$ excitations and a $\sim 20$ meV/$\sim 50$ meV redshift of the ${}^4T_{2\mathrm{g}}$/LMCT peaks, respectively, from the AFM ($B = 0$ T) to FM ($B > 1.8$ T) states. Meanwhile, the MCD spectrum in the FM state of both 2L and 5L [Fig. \ref{fig:fig5}(d)/(e)] are in close relative agreement. 

These results suggest a dependence of the excitation energies on the magnetic state of CrI$_3$. This is non-trivial as it is commonly assumed that MCD at fixed probe energy is proportional to the net magnetization, which is implicitly used in the interpretation of the magnetic ground state. Our observations here for CrI$_3$ are similar to reports for other field-induced AFM to FM transitions in, e.g., Ni$_3$TeO$_6$ \cite{Yokosuk2016}, Ni$_3$V$_2$O$_8$ \cite{Chen2014}, and CrSBr \cite{Wilson2021}. Based on our CTM analysis [Fig. \ref{fig:fig3}(f)], this could be consistent with either a decrease of the charge transfer gap or an increase of the Cr-I hybridization across the field-induced AFM to AFM transition. A modified Cr-I overlap is evidenced by  structural changes across the AFM $\to$ FM transition in 2L CrI$_3$ inferred by Raman spectroscopy \cite{Guo2021} related to the interlayer magneto-structural coupling \cite{Jang2019,Li2020,McGuire2015}. We also observe an increase in the strength of the $dd$ excitations relative to the LMCT exciton in the FM state [Fig. \ref{fig:fig5}(c)/(d)]. In addition to the apparent energetic shift, this suggests that the increased hybridization with ligand states contribute to optically brightening these transitions. While direct field-dependent optical absorption measurements in few-layer CrI$_3$ would be required to definitively support this scenario, we recall that MCD spectroscopy is critical to resolve the overlapping $dd$ and charge transfer peaks near 1.7 eV.

\begin{figure}
\centering

\includegraphics[width = \columnwidth]{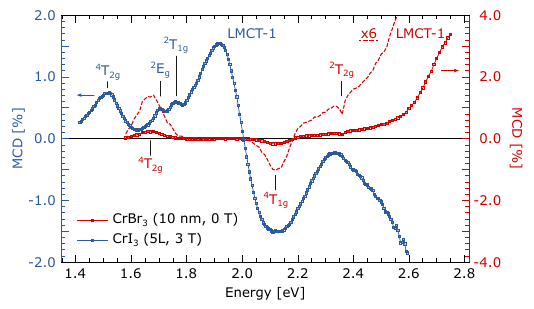}

\caption{Comparison between the MCD spectrum in thin layers of CrI$_3$ (blue, left axis) and CrBr$_3$ (red, right axis) in the fully saturated ferromagnetic state. The dashed red line is multiplied by a factor of 6 for comparison. Individual excitations based on multiplet calculations in Figure \ref{fig:fig3} are indicated for each compound.}

\label{fig:fig6}
\end{figure}

To substantiate the proposed role of metal-ligand covalency, we compare the MCD spectrum of FM 5L CrI$_3$ with that of thin bulk ($\sim 10$ nm thick) FM CrBr$_3$ in Fig. \ref{fig:fig6}. This effectively tunes the covalency through ligand substitution, which is smaller in Br due to the smaller ionic radius. All $dd$ excitations are a factor $> 6$ weaker in CrBr$_3$ -- relative to the primary LMCT transition -- compared to CrI$_3$. Below $2.2$ eV, only spin-allowed transitions are observed and the ${}^2E_g$/${}^2T_{1g}$ peaks, expected at similar energies to CrI$_3$ [Fig. \ref{fig:fig3}(f)] \cite{Bermudez1979i}, are not clearly observed. However, a higher energy spin-forbidden ${}^2T_{2\mathrm{g}}$ transition is observed near 2.33 eV [Fig. \ref{fig:fig3}(f)] \cite{Dillon1966}. Thus, the optical cross section of the intra-configurational $\Delta S = 1$ transitions may be more sensitive to hybridization effects compared to the $\Delta S = 0$ transitions. This could be due to reduced Franck-Condon coupling for the former \cite{Benedek1979}, an alternative brightening mechanism primarily active for inter-configurational excitations \cite{ballhausen1962}. This could also explain why the ${}^2T_{2\mathrm{g}}$ state in CrBr$_3$ is relatively prominent compared to the ${}^2E_\mathrm{g}$/${}^2T_{1\mathrm{g}}$ states, due to its closer energetic proximity to and increased mixing with charge transfer states [Fig. \ref{fig:fig3}(g)].

\section{Discussion}

A consistent interpretation of these observations, including the ligand- and field-dependence of the multiplet energies and their optical brightness, relates to the degree of hybridization between the metal and ligand states. In the ionic limit, $dd$ transitions are parity forbidden and weak. As hybridization and covalency increase, the $dd$ excitations become mixed with the ligand $p$ states. This has the effect of modifying their energies through configuration interactions [Fig. \ref{fig:fig3}(f)] and simultaneously increasing their optical cross section [Fig. \ref{fig:fig5}/\ref{fig:fig6}].

While the excitations of CrI$_3$ can be accurately mapped to Cr$^{3+}$ $dd$ excitations that get modified by charge transfer, the ligand orbitals in real solids form dispersive bands. The large contribution of charge transfer configurations in the case of CrX$_3$ likely signals an importance of these band-like states. Our results are consistent with the conclusions of QS$G\hat{W}$ results, suggesting that ligand covalency in CrX$_3$ is the primary brightening mechanism of the $dd$ excitations \cite{Acharya2022}, though such calculations did not capture the $\Delta S = 1$ excitations of focus here \cite{Acharya2023}. The importance of band-like states is also evidenced by the enhanced photoconductivity at $dd$ transitions in highly covalent systems \cite{Kanazawa1970,Pollini1970,Lebedev2024} and are likely responsible for the momentum dependence of $dd$ excitations \cite{Wang2018}, as recently observed in the nickel dihalides \cite{Occhialini2024} and NiPS$_3$ \cite{He2024}. Notably, the ${}^3A_{2\mathrm{g}} \to {}^1A_{1\mathrm{g}}$ spin-forbidden excitations in these Ni$^{2+}$ systems follow similar trends of optical brightness with ligand covalency \cite{Day1978, Kozielski1972, Banda1986, Kuindersma1981a, Rosseinsky1978, Robbins1976}.

We finally address the linewidths of the spin-forbidden excitations in CrI$_3$. The small linewidths generally observed for spin-forbidden multiplets is a property that has attracted significant interest \cite{Kang2020,Wang2021,Hwangbo2021,Son2022,Occhialini2024,He2024}. For example, the prominent ${}^3A_{2g} \leftrightarrow {}^1A_{1g}$ transition in NiPS$_3$ is nearly resolution limited in both RIXS and optical measurements \cite{Kang2020,Wang2021,Hwangbo2021}. Meanwhile, the ${}^2E_g$/${}^2T_{1g}$ peaks in CrI$_3$ are broader than experimental resolution in both techniques, but remain sharper than the spin-allowed transitions (e.g., ${}^4T_{2g}$). This is a general property related to Franck-Condon effects for which the crucial ingredient is the conservation of the ground state orbital configuration \cite{Lohr1972,Kitzmann2022,Benedek1979}, as also supported by our RIXS and optical measurements.

The increased linewidth of the intra-configurational excitations in CrI$_3$ ($\sim 27$ meV) compared to the Ni$^{2+}$ systems (nearly resolution limited for the ${}^1A_{1\mathrm{g}}$ excitation) could be due to either the higher excitation multiplicity (4/6 for ${}^2E_\mathrm{g}$/${}^2T_{1\mathrm{g}}$, respectively) or a larger contribution of band-like ligand states. Indeed, a distinct linewidth is observed for the ${}^1E_\mathrm{g}$ and ${}^1A_{1\mathrm{g}}$ states in the nickel dihalides, possibly due to their distinct multiplicities \cite{Occhialini2024} which allows an (unresolved) broadening from fine structure in reduced symmetry. In addition, while the ${}^1A_{1\mathrm{g}}$ excitons in NiCl$_2$, NiBr$_2$, and NiPS$_3$ are nearly resolution limited, the corresponding excitation in NiI$_2$ is broader \cite{Son2022,Occhialini2024} similar to the present results in CrI$_3$. One common feature is the proximity to the band gap and the increased hybridization with ligand states which may simultaneously brighten and broaden the excitation. This effect may also be characteristic of the I $5p$ states which are more spatially extended and have larger bandwidth compared to, e.g., Br and S $4p$. 
 
A primary goal for future studies should be to quantitatively describe the microscopic relationship between the optical brightness, excitation linewidth and momentum dependence of $dd$ excitations and their interdependence with ligand electronegativity and bandwidth. Such efforts promise to reveal novel and emergent properties of the excitations in intermediate systems simultaneously exhibiting both strong correlations and covalency.

\section{Conclusion} 

We have reported the observation of ${}^4A_{2\mathrm{g}} \to {}^2E_\mathrm{g}/{}^2T_{1\mathrm{g}}$ spin-forbidden excitations in the MCD and optical spectra of CrI$_3$ which are firmly identified by comparison to Cr $L_3$-edge RIXS and CTM calculations. We showed that such excitations are robust to monolayer limit and revealed a significant field tunability of these excitations with the magnetic state in AFM few-layer CrI$_3$. We associate the optical brightness of these transitions and their modification with magnetic state to the effects of metal-ligand covalency, as supported by our CTM analysis. These results not only clarify the magneto-optical response of the chromium trihalides, but also provide significant new insight into the role of covalency for the mechanisms of field tunable magneto-optical properties and the brightening of spin-forbidden multiplet excitations in correlated insulators.\\

{\it Note added:} During the preparation of this manuscript, we became aware of another RIXS study on CrI$_3$ \cite{He2025}.\\

{\it Acknowledgments.} We acknowledge discussions with Yi Tseng and Shiyu Fan. This material is based upon work supported by the National Science Foundation under Grant No. DMR-2405560. This research was performed in part at the 2-ID (SIX) beamline of the National Synchrotron Light Source II, a U.S. Department of Energy (DOE) Office of Science User Facility operated for the DOE Office of Science by Brookhaven National Laboratory under Contract No. DE-SC0012704.

\appendix

\section{Methods}
\textbf{Sample Preparation.} CrI$_3$ single crystals were grown by chemical vapor transport (CVT) \cite{McGuire2015}. For CrI$_3$, stoichiometric Cr and I$_2$ powders where sealed under vacuum in quartz tubes with diameter of 1.27 cm and length of about 20 cm. The tubes were placed in a two-zone furnace and the temperature was raised with a rate of 2°C/min to 650°C and kept constant for 1 day. After this, the precursors end of the tube was kept constant at 650°C, while the cold end was decreased to a temperature of 600°C in 3 h. The tubes were kept at this temperature gradient for 1 week and then naturally cooled down in the furnace. CrI$_3$ single crystals were found at the cold end of the tubes with dimensions of few mm. 

The single crystals of CrBr$_3$ were also grown by means of the CVT method. Cr and PtBr$_2$ were sealed in an evacuated silica ampule with 20 mm diameter. In order to maintain a pressure of 1 bar at high temperature, an additional 100 mg PtBr$_2$ was taken in a 10 mm length of tube. The entire sealed tube was heated in a two-zone furnace between 625°C and 675°C for a period of one week. As a result of this process, large, shiny flake crystals were grown at the cold side, which are easily exfoliable.

For optical measurements, samples were prepared by mechanical exfoliation of single crystals onto SiO$_2$ (285 nm)/Si substrates. The thickness of few-layer samples was confirmed by a combination of optical contrast and atomic force microscopy (AFMWorkshop). Samples were encapsulated by hBN (thickness $\sim$5-10 nm) and transferred onto sapphire (Al$_2$O$_3$) substrates for optical measurements using a standard dry-transfer technique. All samples were handled inside a $N_2$-filled glovebox to prevent sample degradation. Single crystal bulk samples of CrI$_3$ used for RIXS/XAS measurements were cleaved and transferred into the high vacuum loadlock within an N$_2$-filled environment.

\textbf{Optical measurements.} MCD spectra are recorded in back scattering geometry and measured using a standard photoelastic modulator (PEM, Hind's Instruments) technique by modulating the incident beam polarization between circular left ($\sigma^+$) and right ($\sigma^-$) polarizations [see Fig. \ref{fig:fig1}(a)] with $\lambda/4$ retardance at a frequency $f = 50.1$ kHz. The reflectance differential for $\sigma^+$/$\sigma^-$ at the PEM frequency was measured by a lock-in amplifier (Stanford Research Systems, SR865A). Reflectance measurements were simultaneously recorded using a separate lock-in amplifier (SR830) and a mechanical chopper ($f = 587$ Hz). This reflectance signal is also used as a normalization for the MCD data. We used a supercontinuum laser (NKT Photonics) and a custom monochromator set to provide tunable light between 1.3 and 3.0 eV with a fixed resolution of $\sim$ 5 meV across the full spectral range. A 50 x objective lens (Olympus) was used to focus the beam onto the sample (with spot size $\sim$2-4 $\mu$m, dependent on wavelength). Typical incident powers were between $1$-$100$ $\mu$W. Temperature and magnetic field dependent optical measurements were performed in a Quantum Design Opticool cryostat with a base temperature of $\sim 1.75$ K and in magnetic fields up to 3 T. 

\textbf{RIXS/XAS measurements.} High-resolution RIXS and XAS measurements at the Cr $L_{2,3}$-edges were performed at the 2-ID (SIX) beamline at the National Synchrotron Light Source II, Brookhaven National Laboratory \cite{Dvorak2016}. Measurements were performed at a fixed temperature of 40 K with linear horizontal (LH) polarization. RIXS spectra were recorded with a combined energy resolution of $\Delta E = 27$ meV for most measurements, with higher resolution measurements with $\Delta E = 22$ meV reported in Fig. \ref{fig:fig3}(c), inset. Cr $L_{2,3}$-edge XAS was recorded in partial fluorescence yield (PFY).

\textbf{Cluster calculations.} CrX$_6$ cluster modeling of the RIXS and XAS spectra were performed using charge transfer multiplet (CTM) theory as implemented in the QUANTY software \cite{Haverkort2012, Haverkort2014}. The five Cr $3d$ orbitals and the symmetrized $e_g$ and $t_{2g}$ ligand $np$ orbitals are included in the basis, along with the Cr $2p$ core levels for the calculation of $L_{2,3}$-edge X-ray spectra. 

The parameters in the model include the electronic correlations in the Cr $3d$ shell, parameterized by the Slater integrals $F^2_{dd,i/f}$/$F^4_{dd,i/f}$ for the initial and intermediate RIXS configurations ($2p^63d^3$ and $2p^53d^4$, respectively) and the $2p$-$3d$ multiplet coupling given by the Slater integrals $F^2_{pd}$/$G^1_{pd}$/$G^3_{pd}$. The values for the Slater integrals are taken from Hartree-Fock calculations as provided in Ref. \cite{Haverkort2005}. Atomic values correspond to $80\%$ of these values. A uniform reduction of all Slater integrals from atomic values are taken as a free parameter $\beta$, the nephelauxetic reduction (Table \ref{tab:tab1}). Also included are the atomic spin orbit coupling (SOC) of the $3d$ shell in the initial ($\zeta_{3d,i} = 0.035$ eV) and intermediate ($\zeta_{3d,f} = 0.047$ eV) states, and the $2p$ core level SOC ($\zeta_{2p} = 5.667$ eV). The calculations consider $O_h$ symmetry with the crystal field splitting $10Dq$ taken as a free parameter to fit the data (Table \ref{tab:tab1}). 

We explicitly account for the effects of charge transfer by including the $3d^3$/$3d^4\underline{L}$/$3d^5\underline{L}^2$ configurations in the basis. Accurate calculations of the energy levels in CrX$_3$ compounds are found to necessitate the inclusion of the $3d^5\underline{L}^2$ configuration. The parameters relevant to charge transfer include the $pd$ hybridization for the $e_g$ ($V(e_g)$) and $t_{2g}$ ($V(t_{2g})$) states, the charge transfer gap $\Delta$ and the Coulomb repulsion in the initial ($U_{dd}$) and intermediate ($U_{pd}$) states. Further details on the model and parameters can be found in Refs. \cite{Haverkort2012, DeGroot2008}. While all parameters were taken as variable when optimizing the energy level diagrams, we used parameters from previous XPS measurements on CrBr$_3$ and CrCl$_3$ as starting values \cite{Pollini1989}, namely $U_{dd} = 3.0$ eV, $U_{pd} = 4.5$ eV and $V(e_g) = 1.7$ eV with $V(t_{2g}) = \frac{3}{5} V(e_g)$. While these parameters and the ratio $V(t_{2g})/V(e_g)$ were independently tuned, a satisfactory agreement was found by fixing each of these parameters for all ligands and only varying $\Delta$, $10Dq$ and $\beta$. There is a large cross correlation between many parameters and the optimal fits reported in Table \ref{tab:tab1} should be interpreted accordingly. The energy level diagrams versus ligand are thus calculated through variation of $10Dq$, $\Delta$ and $\beta$, as reported in Table \ref{tab:tab1}. The energy level diagram in Fig. \ref{fig:fig3}(f) includes a covariation of parameters based on linear (quadratic) interpolation of $\beta$ ($10Dq$) versus $\Delta$ from the best fit values for each ligand as reported in Table \ref{tab:tab1}.

For the calculation of the XAS, we used a Lorentzian core hole lifetime broadening of $\Gamma = 400$/$500$ meV for the Cr $L_3$/$L_2$ edges, respectively. The isotropic (polarization averaged) XAS spectrum is reported in Fig. \ref{fig:fig3}(b). For the calculated RIXS map in Fig. \ref{fig:fig3}(e), we used the same lifetime broadening along the incident energy axis and a Lorentzian broadening along the energy transfer axis of 50 meV. An additional Gaussian broadening of 30 meV was applied for $\Delta E < 2.4$ eV and was linearly interpolated up to 500 meV for $2.4 < E < 4.4$ eV. This aims to account for the experimentally-observed broadening of the higher energy excitations at $\Delta E \geq 3.0$ eV. The calculated RIXS map corresponds to the experimental polarization conditions and scattering geometry.


\begin{thebibliography}{93}%
\makeatletter
\providecommand \@ifxundefined [1]{%
 \@ifx{#1\undefined}
}%
\providecommand \@ifnum [1]{%
 \ifnum #1\expandafter \@firstoftwo
 \else \expandafter \@secondoftwo
 \fi
}%
\providecommand \@ifx [1]{%
 \ifx #1\expandafter \@firstoftwo
 \else \expandafter \@secondoftwo
 \fi
}%
\providecommand \natexlab [1]{#1}%
\providecommand \enquote  [1]{``#1''}%
\providecommand \bibnamefont  [1]{#1}%
\providecommand \bibfnamefont [1]{#1}%
\providecommand \citenamefont [1]{#1}%
\providecommand \href@noop [0]{\@secondoftwo}%
\providecommand \href [0]{\begingroup \@sanitize@url \@href}%
\providecommand \@href[1]{\@@startlink{#1}\@@href}%
\providecommand \@@href[1]{\endgroup#1\@@endlink}%
\providecommand \@sanitize@url [0]{\catcode `\\12\catcode `\$12\catcode `\&12\catcode `\#12\catcode `\^12\catcode `\_12\catcode `\%12\relax}%
\providecommand \@@startlink[1]{}%
\providecommand \@@endlink[0]{}%
\providecommand \url  [0]{\begingroup\@sanitize@url \@url }%
\providecommand \@url [1]{\endgroup\@href {#1}{\urlprefix }}%
\providecommand \urlprefix  [0]{URL }%
\providecommand \Eprint [0]{\href }%
\providecommand \doibase [0]{https://doi.org/}%
\providecommand \selectlanguage [0]{\@gobble}%
\providecommand \bibinfo  [0]{\@secondoftwo}%
\providecommand \bibfield  [0]{\@secondoftwo}%
\providecommand \translation [1]{[#1]}%
\providecommand \BibitemOpen [0]{}%
\providecommand \bibitemStop [0]{}%
\providecommand \bibitemNoStop [0]{.\EOS\space}%
\providecommand \EOS [0]{\spacefactor3000\relax}%
\providecommand \BibitemShut  [1]{\csname bibitem#1\endcsname}%
\let\auto@bib@innerbib\@empty
\bibitem [{\citenamefont {Hansen}(1959)}]{Hansen1959}%
  \BibitemOpen
  \bibfield  {author} {\bibinfo {author} {\bibfnamefont {W.~N.}\ \bibnamefont {Hansen}},\ }\bibfield  {title} {\bibinfo {title} {Some magnetic properties of the chromium {(III)} halides at {4.2°K}},\ }\href@noop {} {\bibfield  {journal} {\bibinfo  {journal} {Journal of Applied Physics}\ }\textbf {\bibinfo {volume} {30}},\ \bibinfo {pages} {S304} (\bibinfo {year} {1959})}\BibitemShut {NoStop}%
\bibitem [{\citenamefont {Tsubokawa}(1960)}]{Tsubokawa1960}%
  \BibitemOpen
  \bibfield  {author} {\bibinfo {author} {\bibfnamefont {I.}~\bibnamefont {Tsubokawa}},\ }\bibfield  {title} {\bibinfo {title} {On the magnetic properties of a {CrBr$_3$} single crystal},\ }\href@noop {} {\bibfield  {journal} {\bibinfo  {journal} {Journal of the Physical Society of Japan}\ }\textbf {\bibinfo {volume} {15}},\ \bibinfo {pages} {1664} (\bibinfo {year} {1960})}\BibitemShut {NoStop}%
\bibitem [{\citenamefont {Dillon}\ \emph {et~al.}(1962)\citenamefont {Dillon}, \citenamefont {Kamimura},\ and\ \citenamefont {Remeika}}]{Dillon1962}%
  \BibitemOpen
  \bibfield  {author} {\bibinfo {author} {\bibfnamefont {J.~F.}\ \bibnamefont {Dillon}}, \bibinfo {author} {\bibfnamefont {H.}~\bibnamefont {Kamimura}},\ and\ \bibinfo {author} {\bibfnamefont {J.~P.}\ \bibnamefont {Remeika}},\ }\bibfield  {title} {\bibinfo {title} {Magnetic rotation of visible light by ferromagnet {CrBr$_3$}},\ }\href@noop {} {\bibfield  {journal} {\bibinfo  {journal} {Physical Review Letters}\ }\textbf {\bibinfo {volume} {9}},\ \bibinfo {pages} {161} (\bibinfo {year} {1962})}\BibitemShut {NoStop}%
\bibitem [{\citenamefont {Dillon}\ \emph {et~al.}(1963)\citenamefont {Dillon}, \citenamefont {Kamimura},\ and\ \citenamefont {Remeika}}]{Dillon1963}%
  \BibitemOpen
  \bibfield  {author} {\bibinfo {author} {\bibfnamefont {J.~F.}\ \bibnamefont {Dillon}}, \bibinfo {author} {\bibfnamefont {H.}~\bibnamefont {Kamimura}},\ and\ \bibinfo {author} {\bibfnamefont {J.~P.}\ \bibnamefont {Remeika}},\ }\bibfield  {title} {\bibinfo {title} {Magneto-optical studies of chromium tribromide},\ }\href {https://doi.org/10.1063/1.1729455} {\bibfield  {journal} {\bibinfo  {journal} {Journal of Applied Physics}\ }\textbf {\bibinfo {volume} {34}},\ \bibinfo {pages} {1240} (\bibinfo {year} {1963})}\BibitemShut {NoStop}%
\bibitem [{\citenamefont {Dillon}\ \emph {et~al.}(1966)\citenamefont {Dillon}, \citenamefont {Kamimura},\ and\ \citenamefont {Remeika}}]{Dillon1966}%
  \BibitemOpen
  \bibfield  {author} {\bibinfo {author} {\bibfnamefont {J.~F.}\ \bibnamefont {Dillon}}, \bibinfo {author} {\bibfnamefont {H.}~\bibnamefont {Kamimura}},\ and\ \bibinfo {author} {\bibfnamefont {J.~P.}\ \bibnamefont {Remeika}},\ }\bibfield  {title} {\bibinfo {title} {Magneto-optical properties of ferromagnetic chromium trihalides},\ }\href@noop {} {\bibfield  {journal} {\bibinfo  {journal} {Journal of Physics and Chemistry of Solids}\ }\textbf {\bibinfo {volume} {27}},\ \bibinfo {pages} {1531} (\bibinfo {year} {1966})}\BibitemShut {NoStop}%
\bibitem [{\citenamefont {Krinchik}\ and\ \citenamefont {Chetkin}(1969)}]{Krinchik1969}%
  \BibitemOpen
  \bibfield  {author} {\bibinfo {author} {\bibfnamefont {G.~S.}\ \bibnamefont {Krinchik}}\ and\ \bibinfo {author} {\bibfnamefont {M.~V.}\ \bibnamefont {Chetkin}},\ }\bibfield  {title} {\bibinfo {title} {Transparent ferromagnets},\ }\href@noop {} {\bibfield  {journal} {\bibinfo  {journal} {Soviet Physics Uspekhi}\ }\textbf {\bibinfo {volume} {12}},\ \bibinfo {pages} {307} (\bibinfo {year} {1969})}\BibitemShut {NoStop}%
\bibitem [{\citenamefont {Day}(1978)}]{Day1978}%
  \BibitemOpen
  \bibfield  {author} {\bibinfo {author} {\bibfnamefont {P.}~\bibnamefont {Day}},\ }\bibfield  {title} {\bibinfo {title} {New transparent ferromagnets},\ }\href@noop {} {\bibfield  {journal} {\bibinfo  {journal} {Accounts of Chemical Research}\ }\textbf {\bibinfo {volume} {12}},\ \bibinfo {pages} {236} (\bibinfo {year} {1978})}\BibitemShut {NoStop}%
\bibitem [{\citenamefont {Dillon}\ and\ \citenamefont {Olson}(1965)}]{Dillon1965}%
  \BibitemOpen
  \bibfield  {author} {\bibinfo {author} {\bibfnamefont {J.~F.}\ \bibnamefont {Dillon}}\ and\ \bibinfo {author} {\bibfnamefont {C.~E.}\ \bibnamefont {Olson}},\ }\bibfield  {title} {\bibinfo {title} {Magnetization, resonance, and optical properties of the ferromagnet {CrI$_3$}},\ }\href@noop {} {\bibfield  {journal} {\bibinfo  {journal} {Journal of Applied Physics}\ }\textbf {\bibinfo {volume} {36}},\ \bibinfo {pages} {1259} (\bibinfo {year} {1965})}\BibitemShut {NoStop}%
\bibitem [{\citenamefont {Pedroli}\ \emph {et~al.}(1975)\citenamefont {Pedroli}, \citenamefont {Pollini},\ and\ \citenamefont {Spinolo}}]{Pedroli1975}%
  \BibitemOpen
  \bibfield  {author} {\bibinfo {author} {\bibfnamefont {G.}~\bibnamefont {Pedroli}}, \bibinfo {author} {\bibfnamefont {I.}~\bibnamefont {Pollini}},\ and\ \bibinfo {author} {\bibfnamefont {G.}~\bibnamefont {Spinolo}},\ }\bibfield  {title} {\bibinfo {title} {Specific magnetic rotation spectra and crystal field calculation in {CrBr$_3$} and {CrCl$_3$}},\ }\href@noop {} {\bibfield  {journal} {\bibinfo  {journal} {Journal of Physics C: Solid State Physics}\ }\textbf {\bibinfo {volume} {8}},\ \bibinfo {pages} {2317} (\bibinfo {year} {1975})}\BibitemShut {NoStop}%
\bibitem [{\citenamefont {Carricaburu}\ \emph {et~al.}(1986)\citenamefont {Carricaburu}, \citenamefont {Ferre}, \citenamefont {Mamy}, \citenamefont {Pollini},\ and\ \citenamefont {Thomas}}]{Carricaburu1986}%
  \BibitemOpen
  \bibfield  {author} {\bibinfo {author} {\bibfnamefont {B.}~\bibnamefont {Carricaburu}}, \bibinfo {author} {\bibfnamefont {J.}~\bibnamefont {Ferre}}, \bibinfo {author} {\bibfnamefont {R.}~\bibnamefont {Mamy}}, \bibinfo {author} {\bibfnamefont {I.}~\bibnamefont {Pollini}},\ and\ \bibinfo {author} {\bibfnamefont {J.}~\bibnamefont {Thomas}},\ }\bibfield  {title} {\bibinfo {title} {Optical and electron energy loss experiments in ionic {CrCl$_3$} crystals},\ }\href@noop {} {\bibfield  {journal} {\bibinfo  {journal} {Journal of Physics C: Solid State Physics}\ }\textbf {\bibinfo {volume} {19}},\ \bibinfo {pages} {4985} (\bibinfo {year} {1986})}\BibitemShut {NoStop}%
\bibitem [{\citenamefont {Pollini}\ \emph {et~al.}(1989)\citenamefont {Pollini}, \citenamefont {Thomas}, \citenamefont {Carricaburu},\ and\ \citenamefont {Mamy}}]{Pollini1989}%
  \BibitemOpen
  \bibfield  {author} {\bibinfo {author} {\bibfnamefont {I.}~\bibnamefont {Pollini}}, \bibinfo {author} {\bibfnamefont {J.}~\bibnamefont {Thomas}}, \bibinfo {author} {\bibfnamefont {B.}~\bibnamefont {Carricaburu}},\ and\ \bibinfo {author} {\bibfnamefont {R.}~\bibnamefont {Mamy}},\ }\bibfield  {title} {\bibinfo {title} {Optical and electron energy loss experiments in ionic {CrBr$_3$} crystals},\ }\href@noop {} {\bibfield  {journal} {\bibinfo  {journal} {Journal of Physics: Condensed Matter}\ ,\ \bibinfo {pages} {7695}} (\bibinfo {year} {1989})}\BibitemShut {NoStop}%
\bibitem [{\citenamefont {Bermudez}\ and\ \citenamefont {McClure}(1979{\natexlab{a}})}]{Bermudez1979i}%
  \BibitemOpen
  \bibfield  {author} {\bibinfo {author} {\bibfnamefont {V.~M.}\ \bibnamefont {Bermudez}}\ and\ \bibinfo {author} {\bibfnamefont {D.~S.}\ \bibnamefont {McClure}},\ }\bibfield  {title} {\bibinfo {title} {Spectroscopic studies of the two-dimensional magnetic insulators chromium trichloride and chromium tribromide - {I}},\ }\href@noop {} {\bibfield  {journal} {\bibinfo  {journal} {Journal of Physics and Chemistry of Solids}\ }\textbf {\bibinfo {volume} {40}},\ \bibinfo {pages} {129} (\bibinfo {year} {1979}{\natexlab{a}})}\BibitemShut {NoStop}%
\bibitem [{\citenamefont {Bermudez}\ and\ \citenamefont {McClure}(1979{\natexlab{b}})}]{Bermudez1979ii}%
  \BibitemOpen
  \bibfield  {author} {\bibinfo {author} {\bibfnamefont {V.~M.}\ \bibnamefont {Bermudez}}\ and\ \bibinfo {author} {\bibfnamefont {D.~S.}\ \bibnamefont {McClure}},\ }\bibfield  {title} {\bibinfo {title} {Spectroscopic studies of the two-dimensional magnetic insulators chromium trichloride and chromium tribromide - {II}},\ }\href@noop {} {\bibfield  {journal} {\bibinfo  {journal} {Journal of Physics and Chemistry of Solids}\ }\textbf {\bibinfo {volume} {40}},\ \bibinfo {pages} {149} (\bibinfo {year} {1979}{\natexlab{b}})}\BibitemShut {NoStop}%
\bibitem [{\citenamefont {Grant}\ and\ \citenamefont {Street}(1968)}]{Grant1968}%
  \BibitemOpen
  \bibfield  {author} {\bibinfo {author} {\bibfnamefont {P.~M.}\ \bibnamefont {Grant}}\ and\ \bibinfo {author} {\bibfnamefont {G.~B.}\ \bibnamefont {Street}},\ }\bibfield  {title} {\bibinfo {title} {Optical properties of chromium trihalides in the region 1-11 {eV}},\ }\href@noop {} {\bibfield  {journal} {\bibinfo  {journal} {Bulletin of the American Physical Society}\ } (\bibinfo {year} {1968})}\BibitemShut {NoStop}%
\bibitem [{\citenamefont {Ballhausen}(1962)}]{ballhausen1962}%
  \BibitemOpen
  \bibfield  {author} {\bibinfo {author} {\bibfnamefont {C.~J.}\ \bibnamefont {Ballhausen}},\ }\href@noop {} {\emph {\bibinfo {title} {Introduction to Ligand Field Theory}}}\ (\bibinfo  {publisher} {McGraw-Hill},\ \bibinfo {year} {1962})\BibitemShut {NoStop}%
\bibitem [{\citenamefont {Shinagawa}\ \emph {et~al.}(1996)\citenamefont {Shinagawa}, \citenamefont {Sato}, \citenamefont {Ross}, \citenamefont {McAven},\ and\ \citenamefont {Butler}}]{Shinagawa1996}%
  \BibitemOpen
  \bibfield  {author} {\bibinfo {author} {\bibfnamefont {K.}~\bibnamefont {Shinagawa}}, \bibinfo {author} {\bibfnamefont {H.}~\bibnamefont {Sato}}, \bibinfo {author} {\bibfnamefont {H.~J.}\ \bibnamefont {Ross}}, \bibinfo {author} {\bibfnamefont {L.~F.}\ \bibnamefont {McAven}},\ and\ \bibinfo {author} {\bibfnamefont {P.~H.}\ \bibnamefont {Butler}},\ }\bibfield  {title} {\bibinfo {title} {Charge-transfer transitions in chromium trihalides},\ }\href@noop {} {\bibfield  {journal} {\bibinfo  {journal} {Journal of Physics: Condensed Matter}\ }\textbf {\bibinfo {volume} {8}},\ \bibinfo {pages} {8457} (\bibinfo {year} {1996})}\BibitemShut {NoStop}%
\bibitem [{\citenamefont {McGuire}\ \emph {et~al.}(2015)\citenamefont {McGuire}, \citenamefont {Dixit}, \citenamefont {Cooper},\ and\ \citenamefont {Sales}}]{McGuire2015}%
  \BibitemOpen
  \bibfield  {author} {\bibinfo {author} {\bibfnamefont {M.~A.}\ \bibnamefont {McGuire}}, \bibinfo {author} {\bibfnamefont {H.}~\bibnamefont {Dixit}}, \bibinfo {author} {\bibfnamefont {V.~R.}\ \bibnamefont {Cooper}},\ and\ \bibinfo {author} {\bibfnamefont {B.~C.}\ \bibnamefont {Sales}},\ }\bibfield  {title} {\bibinfo {title} {Coupling of crystal structure and magnetism in the layered, ferromagnetic insulator {CrI$_3$}},\ }\href {https://doi.org/10.1021/cm504242t} {\bibfield  {journal} {\bibinfo  {journal} {Chemistry of Materials}\ }\textbf {\bibinfo {volume} {27}},\ \bibinfo {pages} {612} (\bibinfo {year} {2015})}\BibitemShut {NoStop}%
\bibitem [{\citenamefont {Huang}\ \emph {et~al.}(2017)\citenamefont {Huang}, \citenamefont {Clark}, \citenamefont {Navarro-Moratalla}, \citenamefont {Klein}, \citenamefont {Cheng}, \citenamefont {Seyler}, \citenamefont {Zhong}, \citenamefont {Schmidgall}, \citenamefont {McGuire}, \citenamefont {Cobden}, \citenamefont {Yao}, \citenamefont {Xiao}, \citenamefont {Jarillo-Herrero},\ and\ \citenamefont {Xu}}]{Huang2017b}%
  \BibitemOpen
  \bibfield  {author} {\bibinfo {author} {\bibfnamefont {B.}~\bibnamefont {Huang}}, \bibinfo {author} {\bibfnamefont {G.}~\bibnamefont {Clark}}, \bibinfo {author} {\bibfnamefont {E.}~\bibnamefont {Navarro-Moratalla}}, \bibinfo {author} {\bibfnamefont {D.~R.}\ \bibnamefont {Klein}}, \bibinfo {author} {\bibfnamefont {R.}~\bibnamefont {Cheng}}, \bibinfo {author} {\bibfnamefont {K.~L.}\ \bibnamefont {Seyler}}, \bibinfo {author} {\bibfnamefont {D.}~\bibnamefont {Zhong}}, \bibinfo {author} {\bibfnamefont {E.}~\bibnamefont {Schmidgall}}, \bibinfo {author} {\bibfnamefont {M.~A.}\ \bibnamefont {McGuire}}, \bibinfo {author} {\bibfnamefont {D.~H.}\ \bibnamefont {Cobden}}, \bibinfo {author} {\bibfnamefont {W.}~\bibnamefont {Yao}}, \bibinfo {author} {\bibfnamefont {D.}~\bibnamefont {Xiao}}, \bibinfo {author} {\bibfnamefont {P.}~\bibnamefont {Jarillo-Herrero}},\ and\ \bibinfo {author} {\bibfnamefont {X.}~\bibnamefont {Xu}},\ }\bibfield  {title} {\bibinfo {title} {Layer-dependent ferromagnetism in a {van der Waals} crystal
  down to the monolayer limit},\ }\href@noop {} {\bibfield  {journal} {\bibinfo  {journal} {Nature}\ }\textbf {\bibinfo {volume} {546}},\ \bibinfo {pages} {270} (\bibinfo {year} {2017})}\BibitemShut {NoStop}%
\bibitem [{\citenamefont {Seyler}\ \emph {et~al.}(2018)\citenamefont {Seyler}, \citenamefont {Zhong}, \citenamefont {Klein}, \citenamefont {Gao}, \citenamefont {Zhang}, \citenamefont {Huang}, \citenamefont {Navarro-Moratalla}, \citenamefont {Yang}, \citenamefont {Cobden}, \citenamefont {McGuire}, \citenamefont {Yao}, \citenamefont {Xiao}, \citenamefont {Jarillo-Herrero},\ and\ \citenamefont {Xu}}]{Seyler2018}%
  \BibitemOpen
  \bibfield  {author} {\bibinfo {author} {\bibfnamefont {K.~L.}\ \bibnamefont {Seyler}}, \bibinfo {author} {\bibfnamefont {D.}~\bibnamefont {Zhong}}, \bibinfo {author} {\bibfnamefont {D.~R.}\ \bibnamefont {Klein}}, \bibinfo {author} {\bibfnamefont {S.}~\bibnamefont {Gao}}, \bibinfo {author} {\bibfnamefont {X.}~\bibnamefont {Zhang}}, \bibinfo {author} {\bibfnamefont {B.}~\bibnamefont {Huang}}, \bibinfo {author} {\bibfnamefont {E.}~\bibnamefont {Navarro-Moratalla}}, \bibinfo {author} {\bibfnamefont {L.}~\bibnamefont {Yang}}, \bibinfo {author} {\bibfnamefont {D.~H.}\ \bibnamefont {Cobden}}, \bibinfo {author} {\bibfnamefont {M.~A.}\ \bibnamefont {McGuire}}, \bibinfo {author} {\bibfnamefont {W.}~\bibnamefont {Yao}}, \bibinfo {author} {\bibfnamefont {D.}~\bibnamefont {Xiao}}, \bibinfo {author} {\bibfnamefont {P.}~\bibnamefont {Jarillo-Herrero}},\ and\ \bibinfo {author} {\bibfnamefont {X.}~\bibnamefont {Xu}},\ }\bibfield  {title} {\bibinfo {title} {Ligand-field helical luminescence in a {2D} ferromagnetic
  insulator},\ }\href@noop {} {\bibfield  {journal} {\bibinfo  {journal} {Nature Physics}\ }\textbf {\bibinfo {volume} {14}},\ \bibinfo {pages} {277} (\bibinfo {year} {2018})}\BibitemShut {NoStop}%
\bibitem [{\citenamefont {Song}\ \emph {et~al.}(2021{\natexlab{a}})\citenamefont {Song}, \citenamefont {Anderson}, \citenamefont {Tu}, \citenamefont {Seyler}, \citenamefont {Taniguchi}, \citenamefont {Watanabe}, \citenamefont {McGuire}, \citenamefont {Li}, \citenamefont {Cao}, \citenamefont {Xiao}, \citenamefont {Yao},\ and\ \citenamefont {Xu}}]{Song2021}%
  \BibitemOpen
  \bibfield  {author} {\bibinfo {author} {\bibfnamefont {T.}~\bibnamefont {Song}}, \bibinfo {author} {\bibfnamefont {E.}~\bibnamefont {Anderson}}, \bibinfo {author} {\bibfnamefont {M.~W.-Y.}\ \bibnamefont {Tu}}, \bibinfo {author} {\bibfnamefont {K.}~\bibnamefont {Seyler}}, \bibinfo {author} {\bibfnamefont {T.}~\bibnamefont {Taniguchi}}, \bibinfo {author} {\bibfnamefont {K.}~\bibnamefont {Watanabe}}, \bibinfo {author} {\bibfnamefont {M.~A.}\ \bibnamefont {McGuire}}, \bibinfo {author} {\bibfnamefont {X.}~\bibnamefont {Li}}, \bibinfo {author} {\bibfnamefont {T.}~\bibnamefont {Cao}}, \bibinfo {author} {\bibfnamefont {D.}~\bibnamefont {Xiao}}, \bibinfo {author} {\bibfnamefont {W.}~\bibnamefont {Yao}},\ and\ \bibinfo {author} {\bibfnamefont {X.}~\bibnamefont {Xu}},\ }\bibfield  {title} {\bibinfo {title} {Spin photovoltaic effect in magnetic van der waals heterostructures},\ }\href@noop {} {\bibfield  {journal} {\bibinfo  {journal} {Science Advances}\ }\textbf {\bibinfo {volume} {7}},\ \bibinfo {pages} {eabg8094}
  (\bibinfo {year} {2021}{\natexlab{a}})}\BibitemShut {NoStop}%
\bibitem [{\citenamefont {Siena}\ \emph {et~al.}(2020)\citenamefont {Siena}, \citenamefont {Creutz}, \citenamefont {Regan}, \citenamefont {Malinowski}, \citenamefont {Jiang}, \citenamefont {Kluherz}, \citenamefont {Zhu}, \citenamefont {Lin}, \citenamefont {Yoreo}, \citenamefont {Xu}, \citenamefont {Chu},\ and\ \citenamefont {Gamelin}}]{DeSiena2020}%
  \BibitemOpen
  \bibfield  {author} {\bibinfo {author} {\bibfnamefont {M.~C.~D.}\ \bibnamefont {Siena}}, \bibinfo {author} {\bibfnamefont {S.~E.}\ \bibnamefont {Creutz}}, \bibinfo {author} {\bibfnamefont {A.}~\bibnamefont {Regan}}, \bibinfo {author} {\bibfnamefont {P.}~\bibnamefont {Malinowski}}, \bibinfo {author} {\bibfnamefont {Q.}~\bibnamefont {Jiang}}, \bibinfo {author} {\bibfnamefont {K.~T.}\ \bibnamefont {Kluherz}}, \bibinfo {author} {\bibfnamefont {G.}~\bibnamefont {Zhu}}, \bibinfo {author} {\bibfnamefont {Z.}~\bibnamefont {Lin}}, \bibinfo {author} {\bibfnamefont {J.~J.~D.}\ \bibnamefont {Yoreo}}, \bibinfo {author} {\bibfnamefont {X.}~\bibnamefont {Xu}}, \bibinfo {author} {\bibfnamefont {J.~H.}\ \bibnamefont {Chu}},\ and\ \bibinfo {author} {\bibfnamefont {D.~R.}\ \bibnamefont {Gamelin}},\ }\bibfield  {title} {\bibinfo {title} {Two-dimensional van der waals nanoplatelets with robust ferromagnetism},\ }\href {https://doi.org/10.1021/acs.nanolett.0c00102} {\bibfield  {journal} {\bibinfo  {journal} {Nano Letters}\
  }\textbf {\bibinfo {volume} {20}},\ \bibinfo {pages} {2100} (\bibinfo {year} {2020})}\BibitemShut {NoStop}%
\bibitem [{\citenamefont {Jin}\ \emph {et~al.}(2020)\citenamefont {Jin}, \citenamefont {Kim}, \citenamefont {Ye}, \citenamefont {Ye}, \citenamefont {Rojas}, \citenamefont {Luo}, \citenamefont {Yang}, \citenamefont {Yin}, \citenamefont {Horng}, \citenamefont {Tian}, \citenamefont {Fu}, \citenamefont {Xu}, \citenamefont {Deng}, \citenamefont {Lei}, \citenamefont {Tsen}, \citenamefont {Sun}, \citenamefont {He},\ and\ \citenamefont {Zhao}}]{Jin2020}%
  \BibitemOpen
  \bibfield  {author} {\bibinfo {author} {\bibfnamefont {W.}~\bibnamefont {Jin}}, \bibinfo {author} {\bibfnamefont {H.~H.}\ \bibnamefont {Kim}}, \bibinfo {author} {\bibfnamefont {Z.}~\bibnamefont {Ye}}, \bibinfo {author} {\bibfnamefont {G.}~\bibnamefont {Ye}}, \bibinfo {author} {\bibfnamefont {L.}~\bibnamefont {Rojas}}, \bibinfo {author} {\bibfnamefont {X.}~\bibnamefont {Luo}}, \bibinfo {author} {\bibfnamefont {B.}~\bibnamefont {Yang}}, \bibinfo {author} {\bibfnamefont {F.}~\bibnamefont {Yin}}, \bibinfo {author} {\bibfnamefont {J.~S.~A.}\ \bibnamefont {Horng}}, \bibinfo {author} {\bibfnamefont {S.}~\bibnamefont {Tian}}, \bibinfo {author} {\bibfnamefont {Y.}~\bibnamefont {Fu}}, \bibinfo {author} {\bibfnamefont {G.}~\bibnamefont {Xu}}, \bibinfo {author} {\bibfnamefont {H.}~\bibnamefont {Deng}}, \bibinfo {author} {\bibfnamefont {H.}~\bibnamefont {Lei}}, \bibinfo {author} {\bibfnamefont {A.~W.}\ \bibnamefont {Tsen}}, \bibinfo {author} {\bibfnamefont {K.}~\bibnamefont {Sun}}, \bibinfo {author} {\bibfnamefont
  {R.}~\bibnamefont {He}},\ and\ \bibinfo {author} {\bibfnamefont {L.}~\bibnamefont {Zhao}},\ }\bibfield  {title} {\bibinfo {title} {Observation of the polaronic character of excitons in a two-dimensional semiconducting magnet {CrI$_3$}},\ }\href@noop {} {\bibfield  {journal} {\bibinfo  {journal} {Nature Communications}\ }\textbf {\bibinfo {volume} {11}},\ \bibinfo {pages} {4780} (\bibinfo {year} {2020})}\BibitemShut {NoStop}%
\bibitem [{\citenamefont {Acharya}\ \emph {et~al.}(2022)\citenamefont {Acharya}, \citenamefont {Pashov}, \citenamefont {Rudenko}, \citenamefont {Rösner}, \citenamefont {van Schilfgaarde},\ and\ \citenamefont {Katsnelson}}]{Acharya2022}%
  \BibitemOpen
  \bibfield  {author} {\bibinfo {author} {\bibfnamefont {S.}~\bibnamefont {Acharya}}, \bibinfo {author} {\bibfnamefont {D.}~\bibnamefont {Pashov}}, \bibinfo {author} {\bibfnamefont {A.~N.}\ \bibnamefont {Rudenko}}, \bibinfo {author} {\bibfnamefont {M.}~\bibnamefont {Rösner}}, \bibinfo {author} {\bibfnamefont {M.}~\bibnamefont {van Schilfgaarde}},\ and\ \bibinfo {author} {\bibfnamefont {M.~I.}\ \bibnamefont {Katsnelson}},\ }\bibfield  {title} {\bibinfo {title} {Real- and momentum-space description of the excitons in bulk and monolayer chromium tri-halides},\ }\href@noop {} {\bibfield  {journal} {\bibinfo  {journal} {npj 2D Materials and Applications}\ }\textbf {\bibinfo {volume} {6}},\ \bibinfo {pages} {33} (\bibinfo {year} {2022})}\BibitemShut {NoStop}%
\bibitem [{\citenamefont {Galbiati}\ \emph {et~al.}(2023)\citenamefont {Galbiati}, \citenamefont {Ramiro-Manzano}, \citenamefont {Grau}, \citenamefont {Cantos-Prieto}, \citenamefont {Meseguer-Sánchez}, \citenamefont {Kosic}, \citenamefont {Mione}, \citenamefont {Vilar}, \citenamefont {Cantarero}, \citenamefont {Soriano},\ and\ \citenamefont {Navarro-Moratalla}}]{Galbiati2023}%
  \BibitemOpen
  \bibfield  {author} {\bibinfo {author} {\bibfnamefont {M.}~\bibnamefont {Galbiati}}, \bibinfo {author} {\bibfnamefont {F.}~\bibnamefont {Ramiro-Manzano}}, \bibinfo {author} {\bibfnamefont {J.~J.~P.}\ \bibnamefont {Grau}}, \bibinfo {author} {\bibfnamefont {F.}~\bibnamefont {Cantos-Prieto}}, \bibinfo {author} {\bibfnamefont {J.}~\bibnamefont {Meseguer-Sánchez}}, \bibinfo {author} {\bibfnamefont {I.}~\bibnamefont {Kosic}}, \bibinfo {author} {\bibfnamefont {F.}~\bibnamefont {Mione}}, \bibinfo {author} {\bibfnamefont {A.~P.}\ \bibnamefont {Vilar}}, \bibinfo {author} {\bibfnamefont {A.}~\bibnamefont {Cantarero}}, \bibinfo {author} {\bibfnamefont {D.}~\bibnamefont {Soriano}},\ and\ \bibinfo {author} {\bibfnamefont {E.}~\bibnamefont {Navarro-Moratalla}},\ }\bibfield  {title} {\bibinfo {title} {Monolayer-to-mesoscale modulation of the optical properties in {2D CrI$_3$} mapped by hyperspectral microscopy},\ }\href@noop {} {\bibfield  {journal} {\bibinfo  {journal} {Physical Review Letters}\ }\textbf {\bibinfo
  {volume} {130}},\ \bibinfo {pages} {176901} (\bibinfo {year} {2023})}\BibitemShut {NoStop}%
\bibitem [{\citenamefont {Frisk}\ \emph {et~al.}(2018)\citenamefont {Frisk}, \citenamefont {Duffy}, \citenamefont {Zhang}, \citenamefont {van~der Laan},\ and\ \citenamefont {Hesjedal}}]{Frisk2018}%
  \BibitemOpen
  \bibfield  {author} {\bibinfo {author} {\bibfnamefont {A.}~\bibnamefont {Frisk}}, \bibinfo {author} {\bibfnamefont {L.~B.}\ \bibnamefont {Duffy}}, \bibinfo {author} {\bibfnamefont {S.}~\bibnamefont {Zhang}}, \bibinfo {author} {\bibfnamefont {G.}~\bibnamefont {van~der Laan}},\ and\ \bibinfo {author} {\bibfnamefont {T.}~\bibnamefont {Hesjedal}},\ }\bibfield  {title} {\bibinfo {title} {Magnetic x-ray spectroscopy of two-dimensional {CrI$_3$} layers},\ }\href@noop {} {\bibfield  {journal} {\bibinfo  {journal} {Materials Letters}\ }\textbf {\bibinfo {volume} {232}},\ \bibinfo {pages} {5} (\bibinfo {year} {2018})}\BibitemShut {NoStop}%
\bibitem [{\citenamefont {Choi}\ \emph {et~al.}(2020)\citenamefont {Choi}, \citenamefont {Ryan}, \citenamefont {Haskel}, \citenamefont {McChesney}, \citenamefont {Fabbris}, \citenamefont {McGuire},\ and\ \citenamefont {Kim}}]{Choi2020}%
  \BibitemOpen
  \bibfield  {author} {\bibinfo {author} {\bibfnamefont {Y.}~\bibnamefont {Choi}}, \bibinfo {author} {\bibfnamefont {P.~J.}\ \bibnamefont {Ryan}}, \bibinfo {author} {\bibfnamefont {D.}~\bibnamefont {Haskel}}, \bibinfo {author} {\bibfnamefont {J.~L.}\ \bibnamefont {McChesney}}, \bibinfo {author} {\bibfnamefont {G.}~\bibnamefont {Fabbris}}, \bibinfo {author} {\bibfnamefont {M.~A.}\ \bibnamefont {McGuire}},\ and\ \bibinfo {author} {\bibfnamefont {J.~W.}\ \bibnamefont {Kim}},\ }\bibfield  {title} {\bibinfo {title} {Iodine orbital moment and chromium anisotropy contributions to {CrI$_3$} magnetism},\ }\href@noop {} {\bibfield  {journal} {\bibinfo  {journal} {Applied Physics Letters}\ }\textbf {\bibinfo {volume} {117}},\ \bibinfo {pages} {022411} (\bibinfo {year} {2020})}\BibitemShut {NoStop}%
\bibitem [{\citenamefont {Ghosh}\ \emph {et~al.}(2023)\citenamefont {Ghosh}, \citenamefont {Jönsson}, \citenamefont {Mukkattukavil}, \citenamefont {Kvashnin}, \citenamefont {Phuyal}, \citenamefont {Thunström}, \citenamefont {Agåker}, \citenamefont {Nicolaou}, \citenamefont {Jonak}, \citenamefont {Klingeler}, \citenamefont {Kamalakar}, \citenamefont {Sarkar}, \citenamefont {Vasiliev}, \citenamefont {Butorin}, \citenamefont {Eriksson},\ and\ \citenamefont {Abdel-Hafiez}}]{Ghosh2023}%
  \BibitemOpen
  \bibfield  {author} {\bibinfo {author} {\bibfnamefont {A.}~\bibnamefont {Ghosh}}, \bibinfo {author} {\bibfnamefont {H.~J.~M.}\ \bibnamefont {Jönsson}}, \bibinfo {author} {\bibfnamefont {D.~J.}\ \bibnamefont {Mukkattukavil}}, \bibinfo {author} {\bibfnamefont {Y.}~\bibnamefont {Kvashnin}}, \bibinfo {author} {\bibfnamefont {D.}~\bibnamefont {Phuyal}}, \bibinfo {author} {\bibfnamefont {P.}~\bibnamefont {Thunström}}, \bibinfo {author} {\bibfnamefont {M.}~\bibnamefont {Agåker}}, \bibinfo {author} {\bibfnamefont {A.}~\bibnamefont {Nicolaou}}, \bibinfo {author} {\bibfnamefont {M.}~\bibnamefont {Jonak}}, \bibinfo {author} {\bibfnamefont {R.}~\bibnamefont {Klingeler}}, \bibinfo {author} {\bibfnamefont {M.~V.}\ \bibnamefont {Kamalakar}}, \bibinfo {author} {\bibfnamefont {T.}~\bibnamefont {Sarkar}}, \bibinfo {author} {\bibfnamefont {A.~N.}\ \bibnamefont {Vasiliev}}, \bibinfo {author} {\bibfnamefont {S.~M.}\ \bibnamefont {Butorin}}, \bibinfo {author} {\bibfnamefont {O.}~\bibnamefont {Eriksson}},\ and\ \bibinfo
  {author} {\bibfnamefont {M.}~\bibnamefont {Abdel-Hafiez}},\ }\bibfield  {title} {\bibinfo {title} {Magnetic circular dichroism in the dd excitation in the van der waals magnet {CrI$_3$} probed by resonant inelastic x-ray scattering},\ }\href@noop {} {\bibfield  {journal} {\bibinfo  {journal} {Physical Review B}\ }\textbf {\bibinfo {volume} {107}},\ \bibinfo {pages} {115148} (\bibinfo {year} {2023})}\BibitemShut {NoStop}%
\bibitem [{\citenamefont {Shao}\ \emph {et~al.}(2021)\citenamefont {Shao}, \citenamefont {Karki}, \citenamefont {Huang}, \citenamefont {Feng}, \citenamefont {Sumanasekera}, \citenamefont {Guo}, \citenamefont {Chuang},\ and\ \citenamefont {Freelon}}]{Shao2021}%
  \BibitemOpen
  \bibfield  {author} {\bibinfo {author} {\bibfnamefont {Y.~C.}\ \bibnamefont {Shao}}, \bibinfo {author} {\bibfnamefont {B.}~\bibnamefont {Karki}}, \bibinfo {author} {\bibfnamefont {W.}~\bibnamefont {Huang}}, \bibinfo {author} {\bibfnamefont {X.}~\bibnamefont {Feng}}, \bibinfo {author} {\bibfnamefont {G.}~\bibnamefont {Sumanasekera}}, \bibinfo {author} {\bibfnamefont {J.~H.}\ \bibnamefont {Guo}}, \bibinfo {author} {\bibfnamefont {Y.~D.}\ \bibnamefont {Chuang}},\ and\ \bibinfo {author} {\bibfnamefont {B.}~\bibnamefont {Freelon}},\ }\bibfield  {title} {\bibinfo {title} {Spectroscopic determination of key energy scales for the base hamiltonian of chromium trihalides},\ }\href@noop {} {\bibfield  {journal} {\bibinfo  {journal} {Journal of Physical Chemistry Letters}\ }\textbf {\bibinfo {volume} {12}},\ \bibinfo {pages} {724} (\bibinfo {year} {2021})}\BibitemShut {NoStop}%
\bibitem [{\citenamefont {Zhang}\ \emph {et~al.}(2022)\citenamefont {Zhang}, \citenamefont {Chung}, \citenamefont {Li}, \citenamefont {Wang}, \citenamefont {Wang}, \citenamefont {Huey}, \citenamefont {Yang}, \citenamefont {Goldberger}, \citenamefont {Yao},\ and\ \citenamefont {Zhang}}]{Zhang2022}%
  \BibitemOpen
  \bibfield  {author} {\bibinfo {author} {\bibfnamefont {P.}~\bibnamefont {Zhang}}, \bibinfo {author} {\bibfnamefont {T.~F.}\ \bibnamefont {Chung}}, \bibinfo {author} {\bibfnamefont {Q.}~\bibnamefont {Li}}, \bibinfo {author} {\bibfnamefont {S.}~\bibnamefont {Wang}}, \bibinfo {author} {\bibfnamefont {Q.}~\bibnamefont {Wang}}, \bibinfo {author} {\bibfnamefont {W.~L.}\ \bibnamefont {Huey}}, \bibinfo {author} {\bibfnamefont {S.}~\bibnamefont {Yang}}, \bibinfo {author} {\bibfnamefont {J.~E.}\ \bibnamefont {Goldberger}}, \bibinfo {author} {\bibfnamefont {J.}~\bibnamefont {Yao}},\ and\ \bibinfo {author} {\bibfnamefont {X.}~\bibnamefont {Zhang}},\ }\bibfield  {title} {\bibinfo {title} {All-optical switching of magnetization in atomically thin {CrI$_3$}},\ }\href@noop {} {\bibfield  {journal} {\bibinfo  {journal} {Nature Materials}\ }\textbf {\bibinfo {volume} {21}},\ \bibinfo {pages} {1373} (\bibinfo {year} {2022})}\BibitemShut {NoStop}%
\bibitem [{\citenamefont {Guo}\ \emph {et~al.}(2021)\citenamefont {Guo}, \citenamefont {Jin}, \citenamefont {Ye}, \citenamefont {Ye}, \citenamefont {Xie}, \citenamefont {Yang}, \citenamefont {Kim}, \citenamefont {Yan}, \citenamefont {Fu}, \citenamefont {Tian}, \citenamefont {Lei}, \citenamefont {Tsen}, \citenamefont {Sun}, \citenamefont {Yan}, \citenamefont {He},\ and\ \citenamefont {Zhao}}]{Guo2021}%
  \BibitemOpen
  \bibfield  {author} {\bibinfo {author} {\bibfnamefont {X.}~\bibnamefont {Guo}}, \bibinfo {author} {\bibfnamefont {W.}~\bibnamefont {Jin}}, \bibinfo {author} {\bibfnamefont {Z.}~\bibnamefont {Ye}}, \bibinfo {author} {\bibfnamefont {G.}~\bibnamefont {Ye}}, \bibinfo {author} {\bibfnamefont {H.}~\bibnamefont {Xie}}, \bibinfo {author} {\bibfnamefont {B.}~\bibnamefont {Yang}}, \bibinfo {author} {\bibfnamefont {H.~H.}\ \bibnamefont {Kim}}, \bibinfo {author} {\bibfnamefont {S.}~\bibnamefont {Yan}}, \bibinfo {author} {\bibfnamefont {Y.}~\bibnamefont {Fu}}, \bibinfo {author} {\bibfnamefont {S.}~\bibnamefont {Tian}}, \bibinfo {author} {\bibfnamefont {H.}~\bibnamefont {Lei}}, \bibinfo {author} {\bibfnamefont {A.~W.}\ \bibnamefont {Tsen}}, \bibinfo {author} {\bibfnamefont {K.}~\bibnamefont {Sun}}, \bibinfo {author} {\bibfnamefont {J.~A.}\ \bibnamefont {Yan}}, \bibinfo {author} {\bibfnamefont {R.}~\bibnamefont {He}},\ and\ \bibinfo {author} {\bibfnamefont {L.}~\bibnamefont {Zhao}},\ }\bibfield  {title} {\bibinfo {title}
  {Structural monoclinicity and its coupling to layered magnetism in few-layer {CrI$_3$}},\ }\href@noop {} {\bibfield  {journal} {\bibinfo  {journal} {ACS Nano}\ }\textbf {\bibinfo {volume} {15}},\ \bibinfo {pages} {10444} (\bibinfo {year} {2021})}\BibitemShut {NoStop}%
\bibitem [{\citenamefont {Li}\ \emph {et~al.}(2020)\citenamefont {Li}, \citenamefont {Ye}, \citenamefont {Luo}, \citenamefont {Ye}, \citenamefont {Kim}, \citenamefont {Yang}, \citenamefont {Tian}, \citenamefont {Li}, \citenamefont {Lei}, \citenamefont {Tsen}, \citenamefont {Sun}, \citenamefont {He},\ and\ \citenamefont {Zhao}}]{Li2020}%
  \BibitemOpen
  \bibfield  {author} {\bibinfo {author} {\bibfnamefont {S.}~\bibnamefont {Li}}, \bibinfo {author} {\bibfnamefont {Z.}~\bibnamefont {Ye}}, \bibinfo {author} {\bibfnamefont {X.}~\bibnamefont {Luo}}, \bibinfo {author} {\bibfnamefont {G.}~\bibnamefont {Ye}}, \bibinfo {author} {\bibfnamefont {H.~H.}\ \bibnamefont {Kim}}, \bibinfo {author} {\bibfnamefont {B.}~\bibnamefont {Yang}}, \bibinfo {author} {\bibfnamefont {S.}~\bibnamefont {Tian}}, \bibinfo {author} {\bibfnamefont {C.}~\bibnamefont {Li}}, \bibinfo {author} {\bibfnamefont {H.}~\bibnamefont {Lei}}, \bibinfo {author} {\bibfnamefont {A.~W.}\ \bibnamefont {Tsen}}, \bibinfo {author} {\bibfnamefont {K.}~\bibnamefont {Sun}}, \bibinfo {author} {\bibfnamefont {R.}~\bibnamefont {He}},\ and\ \bibinfo {author} {\bibfnamefont {L.}~\bibnamefont {Zhao}},\ }\bibfield  {title} {\bibinfo {title} {Magnetic-field-induced quantum phase transitions in a van der waals magnet},\ }\href@noop {} {\bibfield  {journal} {\bibinfo  {journal} {Physical Review X}\ }\textbf {\bibinfo
  {volume} {10}},\ \bibinfo {pages} {011075} (\bibinfo {year} {2020})}\BibitemShut {NoStop}%
\bibitem [{\citenamefont {Jang}\ \emph {et~al.}(2019)\citenamefont {Jang}, \citenamefont {Jeong}, \citenamefont {Yoon}, \citenamefont {Ryee},\ and\ \citenamefont {Han}}]{Jang2019}%
  \BibitemOpen
  \bibfield  {author} {\bibinfo {author} {\bibfnamefont {S.~W.}\ \bibnamefont {Jang}}, \bibinfo {author} {\bibfnamefont {M.~Y.}\ \bibnamefont {Jeong}}, \bibinfo {author} {\bibfnamefont {H.}~\bibnamefont {Yoon}}, \bibinfo {author} {\bibfnamefont {S.}~\bibnamefont {Ryee}},\ and\ \bibinfo {author} {\bibfnamefont {M.~J.}\ \bibnamefont {Han}},\ }\bibfield  {title} {\bibinfo {title} {Microscopic understanding of magnetic interactions in bilayer {CrI$_3$}},\ }\href@noop {} {\bibfield  {journal} {\bibinfo  {journal} {Physical Review Materials}\ }\textbf {\bibinfo {volume} {3}},\ \bibinfo {pages} {031001(R)} (\bibinfo {year} {2019})}\BibitemShut {NoStop}%
\bibitem [{\citenamefont {Song}\ \emph {et~al.}(2021{\natexlab{b}})\citenamefont {Song}, \citenamefont {Sun}, \citenamefont {Anderson}, \citenamefont {Wang}, \citenamefont {Qian}, \citenamefont {Taniguchi}, \citenamefont {Watanabe}, \citenamefont {McGuire}, \citenamefont {Stöhr}, \citenamefont {Xiao}, \citenamefont {Cao}, \citenamefont {Wrachtrup},\ and\ \citenamefont {Xu}}]{Song2021b}%
  \BibitemOpen
  \bibfield  {author} {\bibinfo {author} {\bibfnamefont {T.}~\bibnamefont {Song}}, \bibinfo {author} {\bibfnamefont {Q.-C.}\ \bibnamefont {Sun}}, \bibinfo {author} {\bibfnamefont {E.}~\bibnamefont {Anderson}}, \bibinfo {author} {\bibfnamefont {C.}~\bibnamefont {Wang}}, \bibinfo {author} {\bibfnamefont {J.}~\bibnamefont {Qian}}, \bibinfo {author} {\bibfnamefont {T.}~\bibnamefont {Taniguchi}}, \bibinfo {author} {\bibfnamefont {K.}~\bibnamefont {Watanabe}}, \bibinfo {author} {\bibfnamefont {M.~A.}\ \bibnamefont {McGuire}}, \bibinfo {author} {\bibfnamefont {R.}~\bibnamefont {Stöhr}}, \bibinfo {author} {\bibfnamefont {D.}~\bibnamefont {Xiao}}, \bibinfo {author} {\bibfnamefont {T.}~\bibnamefont {Cao}}, \bibinfo {author} {\bibfnamefont {J.}~\bibnamefont {Wrachtrup}},\ and\ \bibinfo {author} {\bibfnamefont {X.}~\bibnamefont {Xu}},\ }\bibfield  {title} {\bibinfo {title} {Direct visualization of magnetic domains and moiré magnetism in twisted {2D} magnets},\ }\href@noop {} {\bibfield  {journal} {\bibinfo  {journal}
  {Science}\ }\textbf {\bibinfo {volume} {374}},\ \bibinfo {pages} {1140} (\bibinfo {year} {2021}{\natexlab{b}})}\BibitemShut {NoStop}%
\bibitem [{\citenamefont {Xu}\ \emph {et~al.}(2022)\citenamefont {Xu}, \citenamefont {Ray}, \citenamefont {Shao}, \citenamefont {Jiang}, \citenamefont {Lee}, \citenamefont {Weber}, \citenamefont {Goldberger}, \citenamefont {Watanabe}, \citenamefont {Taniguchi}, \citenamefont {Muller}, \citenamefont {Mak},\ and\ \citenamefont {Shan}}]{Xu2022}%
  \BibitemOpen
  \bibfield  {author} {\bibinfo {author} {\bibfnamefont {Y.}~\bibnamefont {Xu}}, \bibinfo {author} {\bibfnamefont {A.}~\bibnamefont {Ray}}, \bibinfo {author} {\bibfnamefont {Y.~T.}\ \bibnamefont {Shao}}, \bibinfo {author} {\bibfnamefont {S.}~\bibnamefont {Jiang}}, \bibinfo {author} {\bibfnamefont {K.}~\bibnamefont {Lee}}, \bibinfo {author} {\bibfnamefont {D.}~\bibnamefont {Weber}}, \bibinfo {author} {\bibfnamefont {J.~E.}\ \bibnamefont {Goldberger}}, \bibinfo {author} {\bibfnamefont {K.}~\bibnamefont {Watanabe}}, \bibinfo {author} {\bibfnamefont {T.}~\bibnamefont {Taniguchi}}, \bibinfo {author} {\bibfnamefont {D.~A.}\ \bibnamefont {Muller}}, \bibinfo {author} {\bibfnamefont {K.~F.}\ \bibnamefont {Mak}},\ and\ \bibinfo {author} {\bibfnamefont {J.}~\bibnamefont {Shan}},\ }\bibfield  {title} {\bibinfo {title} {Coexisting ferromagnetic–antiferromagnetic state in twisted bilayer {CrI$_3$}},\ }\href@noop {} {\bibfield  {journal} {\bibinfo  {journal} {Nature Nanotechnology}\ }\textbf {\bibinfo {volume} {17}},\
  \bibinfo {pages} {143} (\bibinfo {year} {2022})}\BibitemShut {NoStop}%
\bibitem [{\citenamefont {Xie}\ \emph {et~al.}(2023)\citenamefont {Xie}, \citenamefont {Luo}, \citenamefont {Ye}, \citenamefont {Sun}, \citenamefont {Ye}, \citenamefont {Sung}, \citenamefont {Ge}, \citenamefont {Yan}, \citenamefont {Fu}, \citenamefont {Tian}, \citenamefont {Lei}, \citenamefont {Sun}, \citenamefont {Hovden}, \citenamefont {He},\ and\ \citenamefont {Zhao}}]{Xie2023}%
  \BibitemOpen
  \bibfield  {author} {\bibinfo {author} {\bibfnamefont {H.}~\bibnamefont {Xie}}, \bibinfo {author} {\bibfnamefont {X.}~\bibnamefont {Luo}}, \bibinfo {author} {\bibfnamefont {Z.}~\bibnamefont {Ye}}, \bibinfo {author} {\bibfnamefont {Z.}~\bibnamefont {Sun}}, \bibinfo {author} {\bibfnamefont {G.}~\bibnamefont {Ye}}, \bibinfo {author} {\bibfnamefont {S.~H.}\ \bibnamefont {Sung}}, \bibinfo {author} {\bibfnamefont {H.}~\bibnamefont {Ge}}, \bibinfo {author} {\bibfnamefont {S.}~\bibnamefont {Yan}}, \bibinfo {author} {\bibfnamefont {Y.}~\bibnamefont {Fu}}, \bibinfo {author} {\bibfnamefont {S.}~\bibnamefont {Tian}}, \bibinfo {author} {\bibfnamefont {H.}~\bibnamefont {Lei}}, \bibinfo {author} {\bibfnamefont {K.}~\bibnamefont {Sun}}, \bibinfo {author} {\bibfnamefont {R.}~\bibnamefont {Hovden}}, \bibinfo {author} {\bibfnamefont {R.}~\bibnamefont {He}},\ and\ \bibinfo {author} {\bibfnamefont {L.}~\bibnamefont {Zhao}},\ }\bibfield  {title} {\bibinfo {title} {Evidence of non-collinear spin texture in magnetic moiré
  superlattices},\ }\href@noop {} {\bibfield  {journal} {\bibinfo  {journal} {Nature Physics}\ }\textbf {\bibinfo {volume} {19}},\ \bibinfo {pages} {1150} (\bibinfo {year} {2023})}\BibitemShut {NoStop}%
\bibitem [{\citenamefont {Wu}\ \emph {et~al.}(2019)\citenamefont {Wu}, \citenamefont {Li}, \citenamefont {Cao},\ and\ \citenamefont {Louie}}]{Wu2019}%
  \BibitemOpen
  \bibfield  {author} {\bibinfo {author} {\bibfnamefont {M.}~\bibnamefont {Wu}}, \bibinfo {author} {\bibfnamefont {Z.}~\bibnamefont {Li}}, \bibinfo {author} {\bibfnamefont {T.}~\bibnamefont {Cao}},\ and\ \bibinfo {author} {\bibfnamefont {S.~G.}\ \bibnamefont {Louie}},\ }\bibfield  {title} {\bibinfo {title} {Physical origin of giant excitonic and magneto-optical responses in two-dimensional ferromagnetic insulators},\ }\href@noop {} {\bibfield  {journal} {\bibinfo  {journal} {Nature Communications}\ }\textbf {\bibinfo {volume} {10}},\ \bibinfo {pages} {2371} (\bibinfo {year} {2019})}\BibitemShut {NoStop}%
\bibitem [{\citenamefont {Molina-Sánchez}\ \emph {et~al.}(2020)\citenamefont {Molina-Sánchez}, \citenamefont {Catarina}, \citenamefont {Sangalli},\ and\ \citenamefont {Fernández-Rossier}}]{Molina2020}%
  \BibitemOpen
  \bibfield  {author} {\bibinfo {author} {\bibfnamefont {A.}~\bibnamefont {Molina-Sánchez}}, \bibinfo {author} {\bibfnamefont {G.}~\bibnamefont {Catarina}}, \bibinfo {author} {\bibfnamefont {D.}~\bibnamefont {Sangalli}},\ and\ \bibinfo {author} {\bibfnamefont {J.}~\bibnamefont {Fernández-Rossier}},\ }\bibfield  {title} {\bibinfo {title} {Magneto-optical response of chromium trihalide monolayers: Chemical trends},\ }\href@noop {} {\bibfield  {journal} {\bibinfo  {journal} {Journal of Materials Chemistry C}\ }\textbf {\bibinfo {volume} {8}},\ \bibinfo {pages} {8856} (\bibinfo {year} {2020})}\BibitemShut {NoStop}%
\bibitem [{\citenamefont {Wu}\ \emph {et~al.}(2022)\citenamefont {Wu}, \citenamefont {Li},\ and\ \citenamefont {Louie}}]{Wu2022}%
  \BibitemOpen
  \bibfield  {author} {\bibinfo {author} {\bibfnamefont {M.}~\bibnamefont {Wu}}, \bibinfo {author} {\bibfnamefont {Z.}~\bibnamefont {Li}},\ and\ \bibinfo {author} {\bibfnamefont {S.~G.}\ \bibnamefont {Louie}},\ }\bibfield  {title} {\bibinfo {title} {Optical and magneto-optical properties of ferromagnetic monolayer {CrBr$_3$}: A first-principles {GW} and {GW plus Bethe-Salpeter} equation study},\ }\href@noop {} {\bibfield  {journal} {\bibinfo  {journal} {Physical Review Materials}\ }\textbf {\bibinfo {volume} {6}},\ \bibinfo {pages} {014008} (\bibinfo {year} {2022})}\BibitemShut {NoStop}%
\bibitem [{\citenamefont {Padmanabhan}\ \emph {et~al.}(2022)\citenamefont {Padmanabhan}, \citenamefont {Buessen}, \citenamefont {Tutchton}, \citenamefont {Kwock}, \citenamefont {Gilinsky}, \citenamefont {Lee}, \citenamefont {McGuire}, \citenamefont {Singamaneni}, \citenamefont {Yarotski}, \citenamefont {Paramekanti}, \citenamefont {Zhu},\ and\ \citenamefont {Prasankumar}}]{Padmanabhan2022}%
  \BibitemOpen
  \bibfield  {author} {\bibinfo {author} {\bibfnamefont {P.}~\bibnamefont {Padmanabhan}}, \bibinfo {author} {\bibfnamefont {F.~L.}\ \bibnamefont {Buessen}}, \bibinfo {author} {\bibfnamefont {R.}~\bibnamefont {Tutchton}}, \bibinfo {author} {\bibfnamefont {K.~W.}\ \bibnamefont {Kwock}}, \bibinfo {author} {\bibfnamefont {S.}~\bibnamefont {Gilinsky}}, \bibinfo {author} {\bibfnamefont {M.~C.}\ \bibnamefont {Lee}}, \bibinfo {author} {\bibfnamefont {M.~A.}\ \bibnamefont {McGuire}}, \bibinfo {author} {\bibfnamefont {S.~R.}\ \bibnamefont {Singamaneni}}, \bibinfo {author} {\bibfnamefont {D.~A.}\ \bibnamefont {Yarotski}}, \bibinfo {author} {\bibfnamefont {A.}~\bibnamefont {Paramekanti}}, \bibinfo {author} {\bibfnamefont {J.~X.}\ \bibnamefont {Zhu}},\ and\ \bibinfo {author} {\bibfnamefont {R.~P.}\ \bibnamefont {Prasankumar}},\ }\bibfield  {title} {\bibinfo {title} {Coherent helicity-dependent spin-phonon oscillations in the ferromagnetic van der waals crystal {CrI$_3$}},\ }\href@noop {} {\bibfield  {journal} {\bibinfo
  {journal} {Nature Communications}\ }\textbf {\bibinfo {volume} {13}},\ \bibinfo {pages} {4473} (\bibinfo {year} {2022})}\BibitemShut {NoStop}%
\bibitem [{\citenamefont {Dabrowski}\ \emph {et~al.}(2022)\citenamefont {Dabrowski}, \citenamefont {Guo}, \citenamefont {Strungaru}, \citenamefont {Keatley}, \citenamefont {Withers}, \citenamefont {Santos},\ and\ \citenamefont {Hicken}}]{Dabrowski2022}%
  \BibitemOpen
  \bibfield  {author} {\bibinfo {author} {\bibfnamefont {M.}~\bibnamefont {Dabrowski}}, \bibinfo {author} {\bibfnamefont {S.}~\bibnamefont {Guo}}, \bibinfo {author} {\bibfnamefont {M.}~\bibnamefont {Strungaru}}, \bibinfo {author} {\bibfnamefont {P.~S.}\ \bibnamefont {Keatley}}, \bibinfo {author} {\bibfnamefont {F.}~\bibnamefont {Withers}}, \bibinfo {author} {\bibfnamefont {E.~J.}\ \bibnamefont {Santos}},\ and\ \bibinfo {author} {\bibfnamefont {R.~J.}\ \bibnamefont {Hicken}},\ }\bibfield  {title} {\bibinfo {title} {All-optical control of spin in a {2D} van der waals magnet},\ }\href@noop {} {\bibfield  {journal} {\bibinfo  {journal} {Nature Communications}\ }\textbf {\bibinfo {volume} {13}},\ \bibinfo {pages} {5976} (\bibinfo {year} {2022})}\BibitemShut {NoStop}%
\bibitem [{\citenamefont {Grzeszczyk}\ \emph {et~al.}(2023)\citenamefont {Grzeszczyk}, \citenamefont {Acharya}, \citenamefont {Pashov}, \citenamefont {Chen}, \citenamefont {Vaklinova}, \citenamefont {van Schilfgaarde}, \citenamefont {Watanabe}, \citenamefont {Taniguchi}, \citenamefont {Novoselov}, \citenamefont {Katsnelson},\ and\ \citenamefont {Koperski}}]{Grzeszczyk2023}%
  \BibitemOpen
  \bibfield  {author} {\bibinfo {author} {\bibfnamefont {M.}~\bibnamefont {Grzeszczyk}}, \bibinfo {author} {\bibfnamefont {S.}~\bibnamefont {Acharya}}, \bibinfo {author} {\bibfnamefont {D.}~\bibnamefont {Pashov}}, \bibinfo {author} {\bibfnamefont {Z.}~\bibnamefont {Chen}}, \bibinfo {author} {\bibfnamefont {K.}~\bibnamefont {Vaklinova}}, \bibinfo {author} {\bibfnamefont {M.}~\bibnamefont {van Schilfgaarde}}, \bibinfo {author} {\bibfnamefont {K.}~\bibnamefont {Watanabe}}, \bibinfo {author} {\bibfnamefont {T.}~\bibnamefont {Taniguchi}}, \bibinfo {author} {\bibfnamefont {K.~S.}\ \bibnamefont {Novoselov}}, \bibinfo {author} {\bibfnamefont {M.~I.}\ \bibnamefont {Katsnelson}},\ and\ \bibinfo {author} {\bibfnamefont {M.}~\bibnamefont {Koperski}},\ }\bibfield  {title} {\bibinfo {title} {Strongly correlated exciton-magnetization system for optical spin pumping in {CrBr$_3$} and {CrI$_3$}},\ }\href@noop {} {\bibfield  {journal} {\bibinfo  {journal} {Advanced Materials}\ }\textbf {\bibinfo {volume} {35}},\ \bibinfo
  {pages} {2209513} (\bibinfo {year} {2023})}\BibitemShut {NoStop}%
\bibitem [{\citenamefont {Gu}\ \emph {et~al.}(2020)\citenamefont {Gu}, \citenamefont {Tan}, \citenamefont {Wan}, \citenamefont {Li}, \citenamefont {Peng}, \citenamefont {Lai}, \citenamefont {Ma}, \citenamefont {Yao}, \citenamefont {Yang}, \citenamefont {Yuan}, \citenamefont {Sun}, \citenamefont {Peng}, \citenamefont {Zhang},\ and\ \citenamefont {Ye}}]{Gu2020}%
  \BibitemOpen
  \bibfield  {author} {\bibinfo {author} {\bibfnamefont {P.}~\bibnamefont {Gu}}, \bibinfo {author} {\bibfnamefont {Q.}~\bibnamefont {Tan}}, \bibinfo {author} {\bibfnamefont {Y.}~\bibnamefont {Wan}}, \bibinfo {author} {\bibfnamefont {Z.}~\bibnamefont {Li}}, \bibinfo {author} {\bibfnamefont {Y.}~\bibnamefont {Peng}}, \bibinfo {author} {\bibfnamefont {J.}~\bibnamefont {Lai}}, \bibinfo {author} {\bibfnamefont {J.}~\bibnamefont {Ma}}, \bibinfo {author} {\bibfnamefont {X.}~\bibnamefont {Yao}}, \bibinfo {author} {\bibfnamefont {S.}~\bibnamefont {Yang}}, \bibinfo {author} {\bibfnamefont {K.}~\bibnamefont {Yuan}}, \bibinfo {author} {\bibfnamefont {D.}~\bibnamefont {Sun}}, \bibinfo {author} {\bibfnamefont {B.}~\bibnamefont {Peng}}, \bibinfo {author} {\bibfnamefont {J.}~\bibnamefont {Zhang}},\ and\ \bibinfo {author} {\bibfnamefont {Y.}~\bibnamefont {Ye}},\ }\bibfield  {title} {\bibinfo {title} {Photoluminescent quantum interference in a van der waals magnet preserved by symmetry breaking},\ }\href@noop {} {\bibfield
  {journal} {\bibinfo  {journal} {ACS Nano}\ }\textbf {\bibinfo {volume} {14}},\ \bibinfo {pages} {1003} (\bibinfo {year} {2020})}\BibitemShut {NoStop}%
\bibitem [{\citenamefont {Kim}\ \emph {et~al.}(2022)\citenamefont {Kim}, \citenamefont {Yoon}, \citenamefont {Ahn}, \citenamefont {Jin}, \citenamefont {Kim}, \citenamefont {Jo}, \citenamefont {Lee}, \citenamefont {Kim},\ and\ \citenamefont {Ryu}}]{Kim2022}%
  \BibitemOpen
  \bibfield  {author} {\bibinfo {author} {\bibfnamefont {S.}~\bibnamefont {Kim}}, \bibinfo {author} {\bibfnamefont {S.}~\bibnamefont {Yoon}}, \bibinfo {author} {\bibfnamefont {H.}~\bibnamefont {Ahn}}, \bibinfo {author} {\bibfnamefont {G.}~\bibnamefont {Jin}}, \bibinfo {author} {\bibfnamefont {H.}~\bibnamefont {Kim}}, \bibinfo {author} {\bibfnamefont {M.~H.}\ \bibnamefont {Jo}}, \bibinfo {author} {\bibfnamefont {C.}~\bibnamefont {Lee}}, \bibinfo {author} {\bibfnamefont {J.}~\bibnamefont {Kim}},\ and\ \bibinfo {author} {\bibfnamefont {S.}~\bibnamefont {Ryu}},\ }\bibfield  {title} {\bibinfo {title} {Photoluminescence path bifurcations by spin flip in two-dimensional {CrPS$_4$}},\ }\href@noop {} {\bibfield  {journal} {\bibinfo  {journal} {ACS Nano}\ }\textbf {\bibinfo {volume} {16}},\ \bibinfo {pages} {16385} (\bibinfo {year} {2022})}\BibitemShut {NoStop}%
\bibitem [{\citenamefont {Multian}\ \emph {et~al.}(2024)\citenamefont {Multian}, \citenamefont {Wu}, \citenamefont {van~der Marel}, \citenamefont {Ubrig},\ and\ \citenamefont {Teyssier}}]{Multian2024}%
  \BibitemOpen
  \bibfield  {author} {\bibinfo {author} {\bibfnamefont {V.}~\bibnamefont {Multian}}, \bibinfo {author} {\bibfnamefont {F.}~\bibnamefont {Wu}}, \bibinfo {author} {\bibfnamefont {D.}~\bibnamefont {van~der Marel}}, \bibinfo {author} {\bibfnamefont {N.}~\bibnamefont {Ubrig}},\ and\ \bibinfo {author} {\bibfnamefont {J.}~\bibnamefont {Teyssier}},\ }\bibfield  {title} {\bibinfo {title} {Brightened optical transition as indicator of multiferroicity in a layered antiferromagnet},\ }\href@noop {} {\bibfield  {journal} {\bibinfo  {journal} {arXiv}\ ,\ \bibinfo {pages} {2405.17144}} (\bibinfo {year} {2024})}\BibitemShut {NoStop}%
\bibitem [{\citenamefont {Sugano}\ and\ \citenamefont {Tsujikawa}(1958)}]{Sugano1958}%
  \BibitemOpen
  \bibfield  {author} {\bibinfo {author} {\bibfnamefont {S.}~\bibnamefont {Sugano}}\ and\ \bibinfo {author} {\bibfnamefont {I.}~\bibnamefont {Tsujikawa}},\ }\bibfield  {title} {\bibinfo {title} {Absorption spectra of {Cr$^{3+}$} in {Al$_2$O$_3$} part {B}. experimental studies of the zeeman effect and other properties of the line spectra},\ }\href@noop {} {\bibfield  {journal} {\bibinfo  {journal} {Journal of the Physical Society of Japan}\ }\textbf {\bibinfo {volume} {13}},\ \bibinfo {pages} {899} (\bibinfo {year} {1958})}\BibitemShut {NoStop}%
\bibitem [{\citenamefont {Liehr}(1963)}]{Liehr1963}%
  \BibitemOpen
  \bibfield  {author} {\bibinfo {author} {\bibfnamefont {A.~D.}\ \bibnamefont {Liehr}},\ }\bibfield  {title} {\bibinfo {title} {Three electron (or hole) cubic ligand field spectrum},\ }\href@noop {} {\bibfield  {journal} {\bibinfo  {journal} {The Journal of Physical Chemistry}\ }\textbf {\bibinfo {volume} {67}},\ \bibinfo {pages} {1314} (\bibinfo {year} {1963})}\BibitemShut {NoStop}%
\bibitem [{\citenamefont {Macfarlane}(1963)}]{Macfarlane1963}%
  \BibitemOpen
  \bibfield  {author} {\bibinfo {author} {\bibfnamefont {R.~M.}\ \bibnamefont {Macfarlane}},\ }\bibfield  {title} {\bibinfo {title} {Analysis of the spectrum of {$d^3$} ions in trigonal crystal fields},\ }\href@noop {} {\bibfield  {journal} {\bibinfo  {journal} {The Journal of Chemical Physics}\ }\textbf {\bibinfo {volume} {39}},\ \bibinfo {pages} {3118} (\bibinfo {year} {1963})}\BibitemShut {NoStop}%
\bibitem [{\citenamefont {Schmidt}\ \emph {et~al.}(2013)\citenamefont {Schmidt}, \citenamefont {Wang}, \citenamefont {Kant}, \citenamefont {Mayr}, \citenamefont {Toth}, \citenamefont {Islam}, \citenamefont {Lake}, \citenamefont {Tsurkan}, \citenamefont {Loidl},\ and\ \citenamefont {Deisenhofer}}]{Schmidt2013}%
  \BibitemOpen
  \bibfield  {author} {\bibinfo {author} {\bibfnamefont {M.}~\bibnamefont {Schmidt}}, \bibinfo {author} {\bibfnamefont {Z.}~\bibnamefont {Wang}}, \bibinfo {author} {\bibfnamefont {C.}~\bibnamefont {Kant}}, \bibinfo {author} {\bibfnamefont {F.}~\bibnamefont {Mayr}}, \bibinfo {author} {\bibfnamefont {S.}~\bibnamefont {Toth}}, \bibinfo {author} {\bibfnamefont {A.~T.}\ \bibnamefont {Islam}}, \bibinfo {author} {\bibfnamefont {B.}~\bibnamefont {Lake}}, \bibinfo {author} {\bibfnamefont {V.}~\bibnamefont {Tsurkan}}, \bibinfo {author} {\bibfnamefont {A.}~\bibnamefont {Loidl}},\ and\ \bibinfo {author} {\bibfnamefont {J.}~\bibnamefont {Deisenhofer}},\ }\bibfield  {title} {\bibinfo {title} {Exciton-magnon transitions in the frustrated chromium antiferromagnets {CuCrO$_2$}, {$\alpha$-CaCr$_2$O$_4$}, {CdCr$_2$O$_4$}, and {ZnCr$_2$O$_4$}},\ }\href@noop {} {\bibfield  {journal} {\bibinfo  {journal} {Physical Review B}\ }\textbf {\bibinfo {volume} {87}},\ \bibinfo {pages} {224424} (\bibinfo {year} {2013})}\BibitemShut
  {NoStop}%
\bibitem [{\citenamefont {Rabia}\ \emph {et~al.}(2014)\citenamefont {Rabia}, \citenamefont {Baldassarre}, \citenamefont {Deisenhofer}, \citenamefont {Tsurkan},\ and\ \citenamefont {Kuntscher}}]{Rabia2014}%
  \BibitemOpen
  \bibfield  {author} {\bibinfo {author} {\bibfnamefont {K.}~\bibnamefont {Rabia}}, \bibinfo {author} {\bibfnamefont {L.}~\bibnamefont {Baldassarre}}, \bibinfo {author} {\bibfnamefont {J.}~\bibnamefont {Deisenhofer}}, \bibinfo {author} {\bibfnamefont {V.}~\bibnamefont {Tsurkan}},\ and\ \bibinfo {author} {\bibfnamefont {C.~A.}\ \bibnamefont {Kuntscher}},\ }\bibfield  {title} {\bibinfo {title} {Evolution of the optical properties of chromium spinels {CdCr$_2$O$_4$}, {HgCr$_2$S$_4$}, and {ZnCr$_2$Se$_4$} under high pressure},\ }\href@noop {} {\bibfield  {journal} {\bibinfo  {journal} {Physical Review B}\ }\textbf {\bibinfo {volume} {89}},\ \bibinfo {pages} {125107} (\bibinfo {year} {2014})}\BibitemShut {NoStop}%
\bibitem [{\citenamefont {Macfarlane}\ and\ \citenamefont {Allen}(1971)}]{Macfarlane1971}%
  \BibitemOpen
  \bibfield  {author} {\bibinfo {author} {\bibfnamefont {R.~M.}\ \bibnamefont {Macfarlane}}\ and\ \bibinfo {author} {\bibfnamefont {J.~W.}\ \bibnamefont {Allen}},\ }\bibfield  {title} {\bibinfo {title} {Exciton bands in antiferromagnetic {Cr$_2$O$_3$}},\ }\href@noop {} {\bibfield  {journal} {\bibinfo  {journal} {Physical Review B}\ }\textbf {\bibinfo {volume} {4}},\ \bibinfo {pages} {3054} (\bibinfo {year} {1971})}\BibitemShut {NoStop}%
\bibitem [{\citenamefont {Banda}(1986)}]{Banda1986}%
  \BibitemOpen
  \bibfield  {author} {\bibinfo {author} {\bibfnamefont {E.~J.}\ \bibnamefont {Banda}},\ }\bibfield  {title} {\bibinfo {title} {Optical absorption of {NiPS$_3$} in the near-infrared, visible and near-ultraviolet regions},\ }\href@noop {} {\bibfield  {journal} {\bibinfo  {journal} {Journal of Physics C: Solid State Physics}\ }\textbf {\bibinfo {volume} {19}},\ \bibinfo {pages} {7329} (\bibinfo {year} {1986})}\BibitemShut {NoStop}%
\bibitem [{\citenamefont {Kang}\ \emph {et~al.}(2020)\citenamefont {Kang}, \citenamefont {Kim}, \citenamefont {Kim}, \citenamefont {Kim}, \citenamefont {Sim}, \citenamefont {Lee}, \citenamefont {Lee}, \citenamefont {Park}, \citenamefont {Yun}, \citenamefont {Kim}, \citenamefont {Nag}, \citenamefont {Walters}, \citenamefont {Garcia-Fernandez}, \citenamefont {Li}, \citenamefont {Chapon}, \citenamefont {Zhou}, \citenamefont {Son}, \citenamefont {Kim}, \citenamefont {Cheong},\ and\ \citenamefont {Park}}]{Kang2020}%
  \BibitemOpen
  \bibfield  {author} {\bibinfo {author} {\bibfnamefont {S.}~\bibnamefont {Kang}}, \bibinfo {author} {\bibfnamefont {K.}~\bibnamefont {Kim}}, \bibinfo {author} {\bibfnamefont {B.~H.}\ \bibnamefont {Kim}}, \bibinfo {author} {\bibfnamefont {J.}~\bibnamefont {Kim}}, \bibinfo {author} {\bibfnamefont {K.~I.}\ \bibnamefont {Sim}}, \bibinfo {author} {\bibfnamefont {J.~U.}\ \bibnamefont {Lee}}, \bibinfo {author} {\bibfnamefont {S.}~\bibnamefont {Lee}}, \bibinfo {author} {\bibfnamefont {K.}~\bibnamefont {Park}}, \bibinfo {author} {\bibfnamefont {S.}~\bibnamefont {Yun}}, \bibinfo {author} {\bibfnamefont {T.}~\bibnamefont {Kim}}, \bibinfo {author} {\bibfnamefont {A.}~\bibnamefont {Nag}}, \bibinfo {author} {\bibfnamefont {A.}~\bibnamefont {Walters}}, \bibinfo {author} {\bibfnamefont {M.}~\bibnamefont {Garcia-Fernandez}}, \bibinfo {author} {\bibfnamefont {J.}~\bibnamefont {Li}}, \bibinfo {author} {\bibfnamefont {L.}~\bibnamefont {Chapon}}, \bibinfo {author} {\bibfnamefont {K.~J.}\ \bibnamefont {Zhou}}, \bibinfo {author}
  {\bibfnamefont {Y.~W.}\ \bibnamefont {Son}}, \bibinfo {author} {\bibfnamefont {J.~H.}\ \bibnamefont {Kim}}, \bibinfo {author} {\bibfnamefont {H.}~\bibnamefont {Cheong}},\ and\ \bibinfo {author} {\bibfnamefont {J.~G.}\ \bibnamefont {Park}},\ }\bibfield  {title} {\bibinfo {title} {Coherent many-body exciton in van der waals antiferromagnet {NiPS$_3$}},\ }\href@noop {} {\bibfield  {journal} {\bibinfo  {journal} {Nature}\ }\textbf {\bibinfo {volume} {583}},\ \bibinfo {pages} {785} (\bibinfo {year} {2020})}\BibitemShut {NoStop}%
\bibitem [{\citenamefont {Hwangbo}\ \emph {et~al.}(2021)\citenamefont {Hwangbo}, \citenamefont {Zhang}, \citenamefont {Jiang}, \citenamefont {Wang}, \citenamefont {Fonseca}, \citenamefont {Wang}, \citenamefont {Diederich}, \citenamefont {Gamelin}, \citenamefont {Xiao}, \citenamefont {Chu}, \citenamefont {Yao},\ and\ \citenamefont {Xu}}]{Hwangbo2021}%
  \BibitemOpen
  \bibfield  {author} {\bibinfo {author} {\bibfnamefont {K.}~\bibnamefont {Hwangbo}}, \bibinfo {author} {\bibfnamefont {Q.}~\bibnamefont {Zhang}}, \bibinfo {author} {\bibfnamefont {Q.}~\bibnamefont {Jiang}}, \bibinfo {author} {\bibfnamefont {Y.}~\bibnamefont {Wang}}, \bibinfo {author} {\bibfnamefont {J.}~\bibnamefont {Fonseca}}, \bibinfo {author} {\bibfnamefont {C.}~\bibnamefont {Wang}}, \bibinfo {author} {\bibfnamefont {G.~M.}\ \bibnamefont {Diederich}}, \bibinfo {author} {\bibfnamefont {D.~R.}\ \bibnamefont {Gamelin}}, \bibinfo {author} {\bibfnamefont {D.}~\bibnamefont {Xiao}}, \bibinfo {author} {\bibfnamefont {J.~H.}\ \bibnamefont {Chu}}, \bibinfo {author} {\bibfnamefont {W.}~\bibnamefont {Yao}},\ and\ \bibinfo {author} {\bibfnamefont {X.}~\bibnamefont {Xu}},\ }\bibfield  {title} {\bibinfo {title} {Highly anisotropic excitons and multiple phonon bound states in a van der waals antiferromagnetic insulator},\ }\href@noop {} {\bibfield  {journal} {\bibinfo  {journal} {Nature Nanotechnology}\ }\textbf
  {\bibinfo {volume} {16}},\ \bibinfo {pages} {655} (\bibinfo {year} {2021})}\BibitemShut {NoStop}%
\bibitem [{\citenamefont {Wang}\ \emph {et~al.}(2021)\citenamefont {Wang}, \citenamefont {Cao}, \citenamefont {Lu}, \citenamefont {Cohen}, \citenamefont {Kitadai}, \citenamefont {Li}, \citenamefont {Tan}, \citenamefont {Wilson}, \citenamefont {Lui}, \citenamefont {Smirnov}, \citenamefont {Sharifzadeh},\ and\ \citenamefont {Ling}}]{Wang2021}%
  \BibitemOpen
  \bibfield  {author} {\bibinfo {author} {\bibfnamefont {X.}~\bibnamefont {Wang}}, \bibinfo {author} {\bibfnamefont {J.}~\bibnamefont {Cao}}, \bibinfo {author} {\bibfnamefont {Z.}~\bibnamefont {Lu}}, \bibinfo {author} {\bibfnamefont {A.}~\bibnamefont {Cohen}}, \bibinfo {author} {\bibfnamefont {H.}~\bibnamefont {Kitadai}}, \bibinfo {author} {\bibfnamefont {T.}~\bibnamefont {Li}}, \bibinfo {author} {\bibfnamefont {Q.}~\bibnamefont {Tan}}, \bibinfo {author} {\bibfnamefont {M.}~\bibnamefont {Wilson}}, \bibinfo {author} {\bibfnamefont {C.~H.}\ \bibnamefont {Lui}}, \bibinfo {author} {\bibfnamefont {D.}~\bibnamefont {Smirnov}}, \bibinfo {author} {\bibfnamefont {S.}~\bibnamefont {Sharifzadeh}},\ and\ \bibinfo {author} {\bibfnamefont {X.}~\bibnamefont {Ling}},\ }\bibfield  {title} {\bibinfo {title} {Spin-induced linear polarization of photoluminescence in antiferromagnetic van der waals crystals},\ }\href@noop {} {\bibfield  {journal} {\bibinfo  {journal} {Nature Materials}\ }\textbf {\bibinfo {volume} {20}},\
  \bibinfo {pages} {964} (\bibinfo {year} {2021})}\BibitemShut {NoStop}%
\bibitem [{\citenamefont {Ergeçen}\ \emph {et~al.}(2022)\citenamefont {Ergeçen}, \citenamefont {Ilyas}, \citenamefont {Mao}, \citenamefont {Po}, \citenamefont {Yilmaz}, \citenamefont {Kim}, \citenamefont {Park}, \citenamefont {Senthil},\ and\ \citenamefont {Gedik}}]{Ergecen2022}%
  \BibitemOpen
  \bibfield  {author} {\bibinfo {author} {\bibfnamefont {E.}~\bibnamefont {Ergeçen}}, \bibinfo {author} {\bibfnamefont {B.}~\bibnamefont {Ilyas}}, \bibinfo {author} {\bibfnamefont {D.}~\bibnamefont {Mao}}, \bibinfo {author} {\bibfnamefont {H.~C.}\ \bibnamefont {Po}}, \bibinfo {author} {\bibfnamefont {M.~B.}\ \bibnamefont {Yilmaz}}, \bibinfo {author} {\bibfnamefont {J.}~\bibnamefont {Kim}}, \bibinfo {author} {\bibfnamefont {J.~G.}\ \bibnamefont {Park}}, \bibinfo {author} {\bibfnamefont {T.}~\bibnamefont {Senthil}},\ and\ \bibinfo {author} {\bibfnamefont {N.}~\bibnamefont {Gedik}},\ }\bibfield  {title} {\bibinfo {title} {Magnetically brightened dark electron-phonon bound states in a van der waals antiferromagnet},\ }\href@noop {} {\bibfield  {journal} {\bibinfo  {journal} {Nature Communications}\ }\textbf {\bibinfo {volume} {13}},\ \bibinfo {pages} {98} (\bibinfo {year} {2022})}\BibitemShut {NoStop}%
\bibitem [{\citenamefont {Jana}\ \emph {et~al.}(2023)\citenamefont {Jana}, \citenamefont {Kapuscinski}, \citenamefont {Mohelsky}, \citenamefont {Vaclavkova}, \citenamefont {Breslavetz}, \citenamefont {Orlita}, \citenamefont {Faugeras},\ and\ \citenamefont {Potemski}}]{Jana2023}%
  \BibitemOpen
  \bibfield  {author} {\bibinfo {author} {\bibfnamefont {D.}~\bibnamefont {Jana}}, \bibinfo {author} {\bibfnamefont {P.}~\bibnamefont {Kapuscinski}}, \bibinfo {author} {\bibfnamefont {I.}~\bibnamefont {Mohelsky}}, \bibinfo {author} {\bibfnamefont {D.}~\bibnamefont {Vaclavkova}}, \bibinfo {author} {\bibfnamefont {I.}~\bibnamefont {Breslavetz}}, \bibinfo {author} {\bibfnamefont {M.}~\bibnamefont {Orlita}}, \bibinfo {author} {\bibfnamefont {C.}~\bibnamefont {Faugeras}},\ and\ \bibinfo {author} {\bibfnamefont {M.}~\bibnamefont {Potemski}},\ }\bibfield  {title} {\bibinfo {title} {Magnon gap excitations and spin-entangled optical transition in the van der waals antiferromagnet {NiPS$_3$}},\ }\href@noop {} {\bibfield  {journal} {\bibinfo  {journal} {Physical Review B}\ }\textbf {\bibinfo {volume} {108}},\ \bibinfo {pages} {115149} (\bibinfo {year} {2023})}\BibitemShut {NoStop}%
\bibitem [{\citenamefont {Kim}\ \emph {et~al.}(2023)\citenamefont {Kim}, \citenamefont {Huang}, \citenamefont {Guo}, \citenamefont {Li}, \citenamefont {Rocca}, \citenamefont {Gao}, \citenamefont {Choe}, \citenamefont {Lujan}, \citenamefont {Wu}, \citenamefont {Lin}, \citenamefont {Baldini}, \citenamefont {Yang}, \citenamefont {Sharma}, \citenamefont {Kalaivanan}, \citenamefont {Sankar}, \citenamefont {Lee}, \citenamefont {Ping},\ and\ \citenamefont {Li}}]{Kim2023}%
  \BibitemOpen
  \bibfield  {author} {\bibinfo {author} {\bibfnamefont {D.~S.}\ \bibnamefont {Kim}}, \bibinfo {author} {\bibfnamefont {D.}~\bibnamefont {Huang}}, \bibinfo {author} {\bibfnamefont {C.}~\bibnamefont {Guo}}, \bibinfo {author} {\bibfnamefont {K.}~\bibnamefont {Li}}, \bibinfo {author} {\bibfnamefont {D.}~\bibnamefont {Rocca}}, \bibinfo {author} {\bibfnamefont {F.~Y.}\ \bibnamefont {Gao}}, \bibinfo {author} {\bibfnamefont {J.}~\bibnamefont {Choe}}, \bibinfo {author} {\bibfnamefont {D.}~\bibnamefont {Lujan}}, \bibinfo {author} {\bibfnamefont {T.~H.}\ \bibnamefont {Wu}}, \bibinfo {author} {\bibfnamefont {K.~H.}\ \bibnamefont {Lin}}, \bibinfo {author} {\bibfnamefont {E.}~\bibnamefont {Baldini}}, \bibinfo {author} {\bibfnamefont {L.}~\bibnamefont {Yang}}, \bibinfo {author} {\bibfnamefont {S.}~\bibnamefont {Sharma}}, \bibinfo {author} {\bibfnamefont {R.}~\bibnamefont {Kalaivanan}}, \bibinfo {author} {\bibfnamefont {R.}~\bibnamefont {Sankar}}, \bibinfo {author} {\bibfnamefont {S.~F.}\ \bibnamefont {Lee}}, \bibinfo
  {author} {\bibfnamefont {Y.}~\bibnamefont {Ping}},\ and\ \bibinfo {author} {\bibfnamefont {X.}~\bibnamefont {Li}},\ }\bibfield  {title} {\bibinfo {title} {Anisotropic excitons reveal local spin chain directions in a van der waals antiferromagnet},\ }\href@noop {} {\bibfield  {journal} {\bibinfo  {journal} {Advanced Materials}\ }\textbf {\bibinfo {volume} {35}},\ \bibinfo {pages} {2206585} (\bibinfo {year} {2023})}\BibitemShut {NoStop}%
\bibitem [{\citenamefont {He}\ \emph {et~al.}(2024)\citenamefont {He}, \citenamefont {Shen}, \citenamefont {Wohlfeld}, \citenamefont {Sears}, \citenamefont {Li}, \citenamefont {Pelliciari}, \citenamefont {Walicki}, \citenamefont {Johnston}, \citenamefont {Baldini}, \citenamefont {Bisogni}, \citenamefont {Mitrano},\ and\ \citenamefont {Dean}}]{He2024}%
  \BibitemOpen
  \bibfield  {author} {\bibinfo {author} {\bibfnamefont {W.}~\bibnamefont {He}}, \bibinfo {author} {\bibfnamefont {Y.}~\bibnamefont {Shen}}, \bibinfo {author} {\bibfnamefont {K.}~\bibnamefont {Wohlfeld}}, \bibinfo {author} {\bibfnamefont {J.}~\bibnamefont {Sears}}, \bibinfo {author} {\bibfnamefont {J.}~\bibnamefont {Li}}, \bibinfo {author} {\bibfnamefont {J.}~\bibnamefont {Pelliciari}}, \bibinfo {author} {\bibfnamefont {M.}~\bibnamefont {Walicki}}, \bibinfo {author} {\bibfnamefont {S.}~\bibnamefont {Johnston}}, \bibinfo {author} {\bibfnamefont {E.}~\bibnamefont {Baldini}}, \bibinfo {author} {\bibfnamefont {V.}~\bibnamefont {Bisogni}}, \bibinfo {author} {\bibfnamefont {M.}~\bibnamefont {Mitrano}},\ and\ \bibinfo {author} {\bibfnamefont {M.~P.}\ \bibnamefont {Dean}},\ }\bibfield  {title} {\bibinfo {title} {Magnetically propagating hund’s exciton in van der waals antiferromagnet {NiPS$_3$}},\ }\href@noop {} {\bibfield  {journal} {\bibinfo  {journal} {Nature Communications}\ }\textbf {\bibinfo {volume} {15}},\
  \bibinfo {pages} {3496} (\bibinfo {year} {2024})}\BibitemShut {NoStop}%
\bibitem [{\citenamefont {Kozielski}\ \emph {et~al.}(1972)\citenamefont {Kozielski}, \citenamefont {Pollini},\ and\ \citenamefont {Spinolo}}]{Kozielski1972}%
  \BibitemOpen
  \bibfield  {author} {\bibinfo {author} {\bibfnamefont {M.}~\bibnamefont {Kozielski}}, \bibinfo {author} {\bibfnamefont {I.}~\bibnamefont {Pollini}},\ and\ \bibinfo {author} {\bibfnamefont {G.}~\bibnamefont {Spinolo}},\ }\bibfield  {title} {\bibinfo {title} {Electronic absorption spectra of {Ni$^{2+}$} in {NiCl$_2$} and {NiBr$_2$}. (phonon and magnon sidebands)},\ }\href@noop {} {\bibfield  {journal} {\bibinfo  {journal} {Journal of Physics C: Solid State Physics}\ }\textbf {\bibinfo {volume} {5}},\ \bibinfo {pages} {1253} (\bibinfo {year} {1972})}\BibitemShut {NoStop}%
\bibitem [{\citenamefont {Giordano}\ \emph {et~al.}(1978)\citenamefont {Giordano}, \citenamefont {Pollini}, \citenamefont {Reattto},\ and\ \citenamefont {Spinolo}}]{Giordano1978}%
  \BibitemOpen
  \bibfield  {author} {\bibinfo {author} {\bibfnamefont {P.}~\bibnamefont {Giordano}}, \bibinfo {author} {\bibfnamefont {I.}~\bibnamefont {Pollini}}, \bibinfo {author} {\bibfnamefont {L.}~\bibnamefont {Reattto}},\ and\ \bibinfo {author} {\bibfnamefont {G.}~\bibnamefont {Spinolo}},\ }\bibfield  {title} {\bibinfo {title} {{${}^1E_g$} spin-forbidden transition in {NiBr$_2$}: temperature dependence},\ }\href@noop {} {\bibfield  {journal} {\bibinfo  {journal} {Physical Review B}\ }\textbf {\bibinfo {volume} {17}},\ \bibinfo {pages} {257} (\bibinfo {year} {1978})}\BibitemShut {NoStop}%
\bibitem [{\citenamefont {Robbins}\ and\ \citenamefont {Day}(1976)}]{Robbins1976}%
  \BibitemOpen
  \bibfield  {author} {\bibinfo {author} {\bibfnamefont {D.~J.}\ \bibnamefont {Robbins}}\ and\ \bibinfo {author} {\bibfnamefont {P.}~\bibnamefont {Day}},\ }\bibfield  {title} {\bibinfo {title} {Temperature variation of exciton-magnon absorption bands in metamagnetic transition-metal dihalides},\ }\href@noop {} {\bibfield  {journal} {\bibinfo  {journal} {Journal of Physics C: Solid State Physics}\ }\textbf {\bibinfo {volume} {9}},\ \bibinfo {pages} {867} (\bibinfo {year} {1976})}\BibitemShut {NoStop}%
\bibitem [{\citenamefont {Rosseinsky}\ and\ \citenamefont {Dorrity}(1978)}]{Rosseinsky1978}%
  \BibitemOpen
  \bibfield  {author} {\bibinfo {author} {\bibfnamefont {D.~R.}\ \bibnamefont {Rosseinsky}}\ and\ \bibinfo {author} {\bibfnamefont {I.~A.}\ \bibnamefont {Dorrity}},\ }\bibfield  {title} {\bibinfo {title} {Absorption spectrum of single crystals of {NiI$_2$} at {300-5 K}},\ }\href@noop {} {\bibfield  {journal} {\bibinfo  {journal} {Inorganic Chemistry}\ }\textbf {\bibinfo {volume} {17}},\ \bibinfo {pages} {1600} (\bibinfo {year} {1978})}\BibitemShut {NoStop}%
\bibitem [{\citenamefont {Son}\ \emph {et~al.}(2022)\citenamefont {Son}, \citenamefont {Lee}, \citenamefont {Kim}, \citenamefont {Kim}, \citenamefont {Kim}, \citenamefont {Na}, \citenamefont {Ju}, \citenamefont {Park}, \citenamefont {Nag}, \citenamefont {Zhou}, \citenamefont {Son}, \citenamefont {Kim}, \citenamefont {Noh}, \citenamefont {Park}, \citenamefont {Lee}, \citenamefont {Cheong}, \citenamefont {Kim},\ and\ \citenamefont {Park}}]{Son2022}%
  \BibitemOpen
  \bibfield  {author} {\bibinfo {author} {\bibfnamefont {S.}~\bibnamefont {Son}}, \bibinfo {author} {\bibfnamefont {Y.}~\bibnamefont {Lee}}, \bibinfo {author} {\bibfnamefont {J.~H.}\ \bibnamefont {Kim}}, \bibinfo {author} {\bibfnamefont {B.~H.}\ \bibnamefont {Kim}}, \bibinfo {author} {\bibfnamefont {C.}~\bibnamefont {Kim}}, \bibinfo {author} {\bibfnamefont {W.}~\bibnamefont {Na}}, \bibinfo {author} {\bibfnamefont {H.}~\bibnamefont {Ju}}, \bibinfo {author} {\bibfnamefont {S.}~\bibnamefont {Park}}, \bibinfo {author} {\bibfnamefont {A.}~\bibnamefont {Nag}}, \bibinfo {author} {\bibfnamefont {K.~J.}\ \bibnamefont {Zhou}}, \bibinfo {author} {\bibfnamefont {Y.~W.}\ \bibnamefont {Son}}, \bibinfo {author} {\bibfnamefont {H.}~\bibnamefont {Kim}}, \bibinfo {author} {\bibfnamefont {W.~S.}\ \bibnamefont {Noh}}, \bibinfo {author} {\bibfnamefont {J.~H.}\ \bibnamefont {Park}}, \bibinfo {author} {\bibfnamefont {J.~S.}\ \bibnamefont {Lee}}, \bibinfo {author} {\bibfnamefont {H.}~\bibnamefont {Cheong}}, \bibinfo {author}
  {\bibfnamefont {J.~H.}\ \bibnamefont {Kim}},\ and\ \bibinfo {author} {\bibfnamefont {J.~G.}\ \bibnamefont {Park}},\ }\bibfield  {title} {\bibinfo {title} {Multiferroic-enabled magnetic-excitons in {2D} quantum-entangled van der waals antiferromagnet {NiI$_2$}},\ }\href@noop {} {\bibfield  {journal} {\bibinfo  {journal} {Advanced Materials}\ }\textbf {\bibinfo {volume} {34}},\ \bibinfo {pages} {2109144} (\bibinfo {year} {2022})}\BibitemShut {NoStop}%
\bibitem [{\citenamefont {Occhialini}\ \emph {et~al.}(2024)\citenamefont {Occhialini}, \citenamefont {Tseng}, \citenamefont {Elnaggar}, \citenamefont {Song}, \citenamefont {Blei}, \citenamefont {Tongay}, \citenamefont {Bisogni}, \citenamefont {Groot}, \citenamefont {Pelliciari},\ and\ \citenamefont {Comin}}]{Occhialini2024}%
  \BibitemOpen
  \bibfield  {author} {\bibinfo {author} {\bibfnamefont {C.~A.}\ \bibnamefont {Occhialini}}, \bibinfo {author} {\bibfnamefont {Y.}~\bibnamefont {Tseng}}, \bibinfo {author} {\bibfnamefont {H.}~\bibnamefont {Elnaggar}}, \bibinfo {author} {\bibfnamefont {Q.}~\bibnamefont {Song}}, \bibinfo {author} {\bibfnamefont {M.}~\bibnamefont {Blei}}, \bibinfo {author} {\bibfnamefont {S.~A.}\ \bibnamefont {Tongay}}, \bibinfo {author} {\bibfnamefont {V.}~\bibnamefont {Bisogni}}, \bibinfo {author} {\bibfnamefont {F.~M.~D.}\ \bibnamefont {Groot}}, \bibinfo {author} {\bibfnamefont {J.}~\bibnamefont {Pelliciari}},\ and\ \bibinfo {author} {\bibfnamefont {R.}~\bibnamefont {Comin}},\ }\bibfield  {title} {\bibinfo {title} {Nature of excitons and their ligand-mediated delocalization in nickel dihalide charge-transfer insulators},\ }\href@noop {} {\bibfield  {journal} {\bibinfo  {journal} {Physical Review X}\ }\textbf {\bibinfo {volume} {14}},\ \bibinfo {pages} {031007} (\bibinfo {year} {2024})}\BibitemShut {NoStop}%
\bibitem [{\citenamefont {Lebedev}\ \emph {et~al.}(2024)\citenamefont {Lebedev}, \citenamefont {Gish}, \citenamefont {Garvey}, \citenamefont {Song}, \citenamefont {Zhou}, \citenamefont {Wang}, \citenamefont {Watanabe}, \citenamefont {Taniguchi}, \citenamefont {Chan}, \citenamefont {Darancet}, \citenamefont {Stern}, \citenamefont {Sangwan},\ and\ \citenamefont {Hersam}}]{Lebedev2024}%
  \BibitemOpen
  \bibfield  {author} {\bibinfo {author} {\bibfnamefont {D.}~\bibnamefont {Lebedev}}, \bibinfo {author} {\bibfnamefont {J.~T.}\ \bibnamefont {Gish}}, \bibinfo {author} {\bibfnamefont {E.~S.}\ \bibnamefont {Garvey}}, \bibinfo {author} {\bibfnamefont {T.~W.}\ \bibnamefont {Song}}, \bibinfo {author} {\bibfnamefont {Q.}~\bibnamefont {Zhou}}, \bibinfo {author} {\bibfnamefont {L.}~\bibnamefont {Wang}}, \bibinfo {author} {\bibfnamefont {K.}~\bibnamefont {Watanabe}}, \bibinfo {author} {\bibfnamefont {T.}~\bibnamefont {Taniguchi}}, \bibinfo {author} {\bibfnamefont {M.~K.}\ \bibnamefont {Chan}}, \bibinfo {author} {\bibfnamefont {P.}~\bibnamefont {Darancet}}, \bibinfo {author} {\bibfnamefont {N.~P.}\ \bibnamefont {Stern}}, \bibinfo {author} {\bibfnamefont {V.~K.}\ \bibnamefont {Sangwan}},\ and\ \bibinfo {author} {\bibfnamefont {M.~C.}\ \bibnamefont {Hersam}},\ }\bibfield  {title} {\bibinfo {title} {Photocurrent spectroscopy of dark magnetic excitons in 2d multiferroic {NiI$_2$}},\ }\href@noop {} {\bibfield  {journal}
  {\bibinfo  {journal} {Advanced Science}\ }\textbf {\bibinfo {volume} {11}},\ \bibinfo {pages} {2407862} (\bibinfo {year} {2024})}\BibitemShut {NoStop}%
\bibitem [{\citenamefont {Sugano}\ \emph {et~al.}(1970)\citenamefont {Sugano}, \citenamefont {Tanabe},\ and\ \citenamefont {Kamimura}}]{Sugano1970}%
  \BibitemOpen
  \bibfield  {author} {\bibinfo {author} {\bibfnamefont {S.}~\bibnamefont {Sugano}}, \bibinfo {author} {\bibfnamefont {Y.}~\bibnamefont {Tanabe}},\ and\ \bibinfo {author} {\bibfnamefont {H.}~\bibnamefont {Kamimura}},\ }\href@noop {} {\emph {\bibinfo {title} {Multiplets of Transition-Metal Ions in Crystals}}}\ (\bibinfo  {publisher} {Academic Press},\ \bibinfo {year} {1970})\BibitemShut {NoStop}%
\bibitem [{\citenamefont {Lohr}(1972)}]{Lohr1972}%
  \BibitemOpen
  \bibfield  {author} {\bibinfo {author} {\bibfnamefont {L.~L.}\ \bibnamefont {Lohr}},\ }\bibfield  {title} {\bibinfo {title} {Spin forbidden electronic excitations in transition metal complexes},\ }\href@noop {} {\bibfield  {journal} {\bibinfo  {journal} {Coordination Chemistry Reviews}\ }\textbf {\bibinfo {volume} {8}},\ \bibinfo {pages} {241} (\bibinfo {year} {1972})}\BibitemShut {NoStop}%
\bibitem [{\citenamefont {Kitzmann}\ \emph {et~al.}(2022)\citenamefont {Kitzmann}, \citenamefont {Moll},\ and\ \citenamefont {Heinze}}]{Kitzmann2022}%
  \BibitemOpen
  \bibfield  {author} {\bibinfo {author} {\bibfnamefont {W.~R.}\ \bibnamefont {Kitzmann}}, \bibinfo {author} {\bibfnamefont {J.}~\bibnamefont {Moll}},\ and\ \bibinfo {author} {\bibfnamefont {K.}~\bibnamefont {Heinze}},\ }\bibfield  {title} {\bibinfo {title} {Spin-flip luminescence},\ }\href@noop {} {\bibfield  {journal} {\bibinfo  {journal} {Photochemical and Photobiological Sciences}\ }\textbf {\bibinfo {volume} {21}},\ \bibinfo {pages} {1309} (\bibinfo {year} {2022})}\BibitemShut {NoStop}%
\bibitem [{\citenamefont {Eremenko}\ and\ \citenamefont {Petrov}(1977)}]{Eremenko1977}%
  \BibitemOpen
  \bibfield  {author} {\bibinfo {author} {\bibfnamefont {V.~V.}\ \bibnamefont {Eremenko}}\ and\ \bibinfo {author} {\bibfnamefont {E.~G.}\ \bibnamefont {Petrov}},\ }\bibfield  {title} {\bibinfo {title} {Light absorption in antiferromagnets},\ }\href@noop {} {\bibfield  {journal} {\bibinfo  {journal} {Advances in Physics}\ }\textbf {\bibinfo {volume} {26}},\ \bibinfo {pages} {31} (\bibinfo {year} {1977})}\BibitemShut {NoStop}%
\bibitem [{\citenamefont {Momma}\ and\ \citenamefont {Izumi}(2008)}]{Momma2008}%
  \BibitemOpen
  \bibfield  {author} {\bibinfo {author} {\bibfnamefont {K.}~\bibnamefont {Momma}}\ and\ \bibinfo {author} {\bibfnamefont {F.}~\bibnamefont {Izumi}},\ }\bibfield  {title} {\bibinfo {title} {Vesta: A three-dimensional visualization system for electronic and structural analysis},\ }\href@noop {} {\bibfield  {journal} {\bibinfo  {journal} {Journal of Applied Crystallography}\ }\textbf {\bibinfo {volume} {41}},\ \bibinfo {pages} {653} (\bibinfo {year} {2008})}\BibitemShut {NoStop}%
\bibitem [{\citenamefont {de~Groot}\ and\ \citenamefont {Kotani}(2008)}]{DeGroot2008}%
  \BibitemOpen
  \bibfield  {author} {\bibinfo {author} {\bibfnamefont {F.}~\bibnamefont {de~Groot}}\ and\ \bibinfo {author} {\bibfnamefont {A.}~\bibnamefont {Kotani}},\ }\href@noop {} {\emph {\bibinfo {title} {Core level spectroscopy of solids}}}\ (\bibinfo  {publisher} {CRC Press},\ \bibinfo {year} {2008})\ pp.\ \bibinfo {pages} {1--491}\BibitemShut {NoStop}%
\bibitem [{\citenamefont {Cowan}(1981)}]{Cowan1981}%
  \BibitemOpen
  \bibfield  {author} {\bibinfo {author} {\bibfnamefont {R.~D.}\ \bibnamefont {Cowan}},\ }\href@noop {} {\emph {\bibinfo {title} {The Theory of Atomic Structure and Spectra}}}\ (\bibinfo  {publisher} {University of California Press},\ \bibinfo {year} {1981})\BibitemShut {NoStop}%
\bibitem [{\citenamefont {Haverkort}\ \emph {et~al.}(2012)\citenamefont {Haverkort}, \citenamefont {Zwierzycki},\ and\ \citenamefont {Andersen}}]{Haverkort2012}%
  \BibitemOpen
  \bibfield  {author} {\bibinfo {author} {\bibfnamefont {M.~W.}\ \bibnamefont {Haverkort}}, \bibinfo {author} {\bibfnamefont {M.}~\bibnamefont {Zwierzycki}},\ and\ \bibinfo {author} {\bibfnamefont {O.~K.}\ \bibnamefont {Andersen}},\ }\bibfield  {title} {\bibinfo {title} {Multiplet ligand-field theory using wannier orbitals},\ }\href@noop {} {\bibfield  {journal} {\bibinfo  {journal} {Physical Review B}\ }\textbf {\bibinfo {volume} {85}},\ \bibinfo {pages} {165113} (\bibinfo {year} {2012})}\BibitemShut {NoStop}%
\bibitem [{\citenamefont {Dvorak}\ \emph {et~al.}(2016)\citenamefont {Dvorak}, \citenamefont {Jarrige}, \citenamefont {Bisogni}, \citenamefont {Coburn},\ and\ \citenamefont {Leonhardt}}]{Dvorak2016}%
  \BibitemOpen
  \bibfield  {author} {\bibinfo {author} {\bibfnamefont {J.}~\bibnamefont {Dvorak}}, \bibinfo {author} {\bibfnamefont {I.}~\bibnamefont {Jarrige}}, \bibinfo {author} {\bibfnamefont {V.}~\bibnamefont {Bisogni}}, \bibinfo {author} {\bibfnamefont {S.}~\bibnamefont {Coburn}},\ and\ \bibinfo {author} {\bibfnamefont {W.}~\bibnamefont {Leonhardt}},\ }\bibfield  {title} {\bibinfo {title} {Towards 10 {meV} resolution: {The} design of an ultrahigh resolution soft {X-ray} rixs spectrometer},\ }\href@noop {} {\bibfield  {journal} {\bibinfo  {journal} {Review of Scientific Instruments}\ }\textbf {\bibinfo {volume} {87}},\ \bibinfo {pages} {115109} (\bibinfo {year} {2016})}\BibitemShut {NoStop}%
\bibitem [{\citenamefont {Haverkort}\ \emph {et~al.}(2014)\citenamefont {Haverkort}, \citenamefont {Sangiovanni}, \citenamefont {Hansmann}, \citenamefont {Toschi}, \citenamefont {Lu},\ and\ \citenamefont {Macke}}]{Haverkort2014}%
  \BibitemOpen
  \bibfield  {author} {\bibinfo {author} {\bibfnamefont {M.~W.}\ \bibnamefont {Haverkort}}, \bibinfo {author} {\bibfnamefont {G.}~\bibnamefont {Sangiovanni}}, \bibinfo {author} {\bibfnamefont {P.}~\bibnamefont {Hansmann}}, \bibinfo {author} {\bibfnamefont {A.}~\bibnamefont {Toschi}}, \bibinfo {author} {\bibfnamefont {Y.}~\bibnamefont {Lu}},\ and\ \bibinfo {author} {\bibfnamefont {S.}~\bibnamefont {Macke}},\ }\bibfield  {title} {\bibinfo {title} {Bands, resonances, edge singularities and excitons in core level spectroscopy investigated within the dynamical mean-field theory},\ }\href@noop {} {\bibfield  {journal} {\bibinfo  {journal} {EPL}\ }\textbf {\bibinfo {volume} {108}},\ \bibinfo {pages} {57004} (\bibinfo {year} {2014})}\BibitemShut {NoStop}%
\bibitem [{\citenamefont {Pollini}(1999)}]{Pollini1999}%
  \BibitemOpen
  \bibfield  {author} {\bibinfo {author} {\bibfnamefont {I.}~\bibnamefont {Pollini}},\ }\bibfield  {title} {\bibinfo {title} {Electronic structure of {CrBr$_3$} studied by x-ray photoelectron spectroscopy},\ }\href@noop {} {\bibfield  {journal} {\bibinfo  {journal} {Physical Review B}\ ,\ \bibinfo {pages} {16170}} (\bibinfo {year} {1999})}\BibitemShut {NoStop}%
\bibitem [{\citenamefont {Zaanen}\ \emph {et~al.}(1985)\citenamefont {Zaanen}, \citenamefont {Sawatzky},\ and\ \citenamefont {Allen}}]{Zaanen1985}%
  \BibitemOpen
  \bibfield  {author} {\bibinfo {author} {\bibfnamefont {J.}~\bibnamefont {Zaanen}}, \bibinfo {author} {\bibfnamefont {G.~A.}\ \bibnamefont {Sawatzky}},\ and\ \bibinfo {author} {\bibfnamefont {J.~W.}\ \bibnamefont {Allen}},\ }\bibfield  {title} {\bibinfo {title} {Band gaps and electronic structure of transition-metal compounds},\ }\href@noop {} {\bibfield  {journal} {\bibinfo  {journal} {Physical Review Letters}\ }\textbf {\bibinfo {volume} {55}},\ \bibinfo {pages} {418} (\bibinfo {year} {1985})}\BibitemShut {NoStop}%
\bibitem [{\citenamefont {Wang}\ \emph {et~al.}(2022)\citenamefont {Wang}, \citenamefont {Cao}, \citenamefont {Li}, \citenamefont {Lu}, \citenamefont {Cohen}, \citenamefont {Haldar}, \citenamefont {Kitadai}, \citenamefont {Tan}, \citenamefont {Burch}, \citenamefont {Smirnov}, \citenamefont {Xu}, \citenamefont {Sharifzadeh}, \citenamefont {Liang},\ and\ \citenamefont {Ling}}]{Wang2022}%
  \BibitemOpen
  \bibfield  {author} {\bibinfo {author} {\bibfnamefont {X.}~\bibnamefont {Wang}}, \bibinfo {author} {\bibfnamefont {J.}~\bibnamefont {Cao}}, \bibinfo {author} {\bibfnamefont {H.}~\bibnamefont {Li}}, \bibinfo {author} {\bibfnamefont {Z.}~\bibnamefont {Lu}}, \bibinfo {author} {\bibfnamefont {A.}~\bibnamefont {Cohen}}, \bibinfo {author} {\bibfnamefont {A.}~\bibnamefont {Haldar}}, \bibinfo {author} {\bibfnamefont {H.}~\bibnamefont {Kitadai}}, \bibinfo {author} {\bibfnamefont {Q.}~\bibnamefont {Tan}}, \bibinfo {author} {\bibfnamefont {K.~S.}\ \bibnamefont {Burch}}, \bibinfo {author} {\bibfnamefont {D.}~\bibnamefont {Smirnov}}, \bibinfo {author} {\bibfnamefont {W.}~\bibnamefont {Xu}}, \bibinfo {author} {\bibfnamefont {S.}~\bibnamefont {Sharifzadeh}}, \bibinfo {author} {\bibfnamefont {L.}~\bibnamefont {Liang}},\ and\ \bibinfo {author} {\bibfnamefont {X.}~\bibnamefont {Ling}},\ }\bibfield  {title} {\bibinfo {title} {Electronic raman scattering in the {2D} antiferromagnet {NiPS$_3$}},\ }\href@noop {} {\bibfield
  {journal} {\bibinfo  {journal} {Science Advances}\ }\textbf {\bibinfo {volume} {8}},\ \bibinfo {pages} {7707} (\bibinfo {year} {2022})}\BibitemShut {NoStop}%
\bibitem [{\citenamefont {DiScala}\ \emph {et~al.}(2024)\citenamefont {DiScala}, \citenamefont {Staros}, \citenamefont {de~la Torre}, \citenamefont {Lopez}, \citenamefont {Wong}, \citenamefont {Schulz}, \citenamefont {Barkowiak}, \citenamefont {Bisogni}, \citenamefont {Pelliciari}, \citenamefont {Rubenstein},\ and\ \citenamefont {Plumb}}]{DiScala2024}%
  \BibitemOpen
  \bibfield  {author} {\bibinfo {author} {\bibfnamefont {M.~F.}\ \bibnamefont {DiScala}}, \bibinfo {author} {\bibfnamefont {D.}~\bibnamefont {Staros}}, \bibinfo {author} {\bibfnamefont {A.}~\bibnamefont {de~la Torre}}, \bibinfo {author} {\bibfnamefont {A.}~\bibnamefont {Lopez}}, \bibinfo {author} {\bibfnamefont {D.}~\bibnamefont {Wong}}, \bibinfo {author} {\bibfnamefont {C.}~\bibnamefont {Schulz}}, \bibinfo {author} {\bibfnamefont {M.}~\bibnamefont {Barkowiak}}, \bibinfo {author} {\bibfnamefont {V.}~\bibnamefont {Bisogni}}, \bibinfo {author} {\bibfnamefont {J.}~\bibnamefont {Pelliciari}}, \bibinfo {author} {\bibfnamefont {B.}~\bibnamefont {Rubenstein}},\ and\ \bibinfo {author} {\bibfnamefont {K.~W.}\ \bibnamefont {Plumb}},\ }\bibfield  {title} {\bibinfo {title} {Elucidating the role of dimensionality on the electronic structure of the van der waals antiferromagnet {NiPS$_3$}},\ }\href@noop {} {\bibfield  {journal} {\bibinfo  {journal} {Advanced Physics Research}\ }\textbf {\bibinfo {volume} {3}},\ \bibinfo
  {pages} {2300096} (\bibinfo {year} {2024})}\BibitemShut {NoStop}%
\bibitem [{\citenamefont {Benedek}\ \emph {et~al.}(1979)\citenamefont {Benedek}, \citenamefont {Pollini}, \citenamefont {Piseri},\ and\ \citenamefont {Tubino}}]{Benedek1979}%
  \BibitemOpen
  \bibfield  {author} {\bibinfo {author} {\bibfnamefont {G.}~\bibnamefont {Benedek}}, \bibinfo {author} {\bibfnamefont {I.}~\bibnamefont {Pollini}}, \bibinfo {author} {\bibfnamefont {L.}~\bibnamefont {Piseri}},\ and\ \bibinfo {author} {\bibfnamefont {R.}~\bibnamefont {Tubino}},\ }\bibfield  {title} {\bibinfo {title} {Evidence of two-phonon vibronic progressions in layered 3d-metal dihalides},\ }\href@noop {} {\bibfield  {journal} {\bibinfo  {journal} {Physical Review B}\ }\textbf {\bibinfo {volume} {20}},\ \bibinfo {pages} {4303} (\bibinfo {year} {1979})}\BibitemShut {NoStop}%
\bibitem [{\citenamefont {Chen}\ \emph {et~al.}(2018)\citenamefont {Chen}, \citenamefont {Chung}, \citenamefont {Gao}, \citenamefont {Chen}, \citenamefont {Stone}, \citenamefont {Kolesnikov}, \citenamefont {Huang},\ and\ \citenamefont {Dai}}]{Chen2018}%
  \BibitemOpen
  \bibfield  {author} {\bibinfo {author} {\bibfnamefont {L.}~\bibnamefont {Chen}}, \bibinfo {author} {\bibfnamefont {J.~H.}\ \bibnamefont {Chung}}, \bibinfo {author} {\bibfnamefont {B.}~\bibnamefont {Gao}}, \bibinfo {author} {\bibfnamefont {T.}~\bibnamefont {Chen}}, \bibinfo {author} {\bibfnamefont {M.~B.}\ \bibnamefont {Stone}}, \bibinfo {author} {\bibfnamefont {A.~I.}\ \bibnamefont {Kolesnikov}}, \bibinfo {author} {\bibfnamefont {Q.}~\bibnamefont {Huang}},\ and\ \bibinfo {author} {\bibfnamefont {P.}~\bibnamefont {Dai}},\ }\bibfield  {title} {\bibinfo {title} {Topological spin excitations in honeycomb ferromagnet {CrI$_3$}},\ }\href@noop {} {\bibfield  {journal} {\bibinfo  {journal} {Physical Review X}\ }\textbf {\bibinfo {volume} {8}},\ \bibinfo {pages} {041028} (\bibinfo {year} {2018})}\BibitemShut {NoStop}%
\bibitem [{\citenamefont {Djurdjić-Mijin}\ \emph {et~al.}(2018)\citenamefont {Djurdjić-Mijin}, \citenamefont {Šolajić}, \citenamefont {Pešić}, \citenamefont {Šćepanović}, \citenamefont {Liu}, \citenamefont {Baum}, \citenamefont {Petrovic}, \citenamefont {Lazarević},\ and\ \citenamefont {Popović}}]{Djurdjic2018}%
  \BibitemOpen
  \bibfield  {author} {\bibinfo {author} {\bibfnamefont {S.}~\bibnamefont {Djurdjić-Mijin}}, \bibinfo {author} {\bibfnamefont {A.}~\bibnamefont {Šolajić}}, \bibinfo {author} {\bibfnamefont {J.}~\bibnamefont {Pešić}}, \bibinfo {author} {\bibfnamefont {M.}~\bibnamefont {Šćepanović}}, \bibinfo {author} {\bibfnamefont {Y.}~\bibnamefont {Liu}}, \bibinfo {author} {\bibfnamefont {A.}~\bibnamefont {Baum}}, \bibinfo {author} {\bibfnamefont {C.}~\bibnamefont {Petrovic}}, \bibinfo {author} {\bibfnamefont {N.}~\bibnamefont {Lazarević}},\ and\ \bibinfo {author} {\bibfnamefont {Z.~V.}\ \bibnamefont {Popović}},\ }\bibfield  {title} {\bibinfo {title} {Lattice dynamics and phase transition in {CrI$_3$} single crystals},\ }\href@noop {} {\bibfield  {journal} {\bibinfo  {journal} {Physical Review B}\ }\textbf {\bibinfo {volume} {98}},\ \bibinfo {pages} {104307} (\bibinfo {year} {2018})}\BibitemShut {NoStop}%
\bibitem [{\citenamefont {Song}\ \emph {et~al.}(2018)\citenamefont {Song}, \citenamefont {Cai}, \citenamefont {Tu}, \citenamefont {Zhang}, \citenamefont {Huang}, \citenamefont {Wilson}, \citenamefont {Seyler}, \citenamefont {Zhu}, \citenamefont {Taniguchi}, \citenamefont {Watanabe}, \citenamefont {McGuire}, \citenamefont {Cobden}, \citenamefont {Xiao}, \citenamefont {Yao},\ and\ \citenamefont {Xu}}]{Song2018}%
  \BibitemOpen
  \bibfield  {author} {\bibinfo {author} {\bibfnamefont {T.}~\bibnamefont {Song}}, \bibinfo {author} {\bibfnamefont {X.}~\bibnamefont {Cai}}, \bibinfo {author} {\bibfnamefont {M.~W.~Y.}\ \bibnamefont {Tu}}, \bibinfo {author} {\bibfnamefont {X.}~\bibnamefont {Zhang}}, \bibinfo {author} {\bibfnamefont {B.}~\bibnamefont {Huang}}, \bibinfo {author} {\bibfnamefont {N.~P.}\ \bibnamefont {Wilson}}, \bibinfo {author} {\bibfnamefont {K.~L.}\ \bibnamefont {Seyler}}, \bibinfo {author} {\bibfnamefont {L.}~\bibnamefont {Zhu}}, \bibinfo {author} {\bibfnamefont {T.}~\bibnamefont {Taniguchi}}, \bibinfo {author} {\bibfnamefont {K.}~\bibnamefont {Watanabe}}, \bibinfo {author} {\bibfnamefont {M.~A.}\ \bibnamefont {McGuire}}, \bibinfo {author} {\bibfnamefont {D.~H.}\ \bibnamefont {Cobden}}, \bibinfo {author} {\bibfnamefont {D.}~\bibnamefont {Xiao}}, \bibinfo {author} {\bibfnamefont {W.}~\bibnamefont {Yao}},\ and\ \bibinfo {author} {\bibfnamefont {X.}~\bibnamefont {Xu}},\ }\bibfield  {title} {\bibinfo {title} {Giant tunneling
  magnetoresistance in spin-filter van der waals heterostructures},\ }\href@noop {} {\bibfield  {journal} {\bibinfo  {journal} {Science}\ }\textbf {\bibinfo {volume} {360}},\ \bibinfo {pages} {1214} (\bibinfo {year} {2018})}\BibitemShut {NoStop}%
\bibitem [{\citenamefont {Yokosuk}\ \emph {et~al.}(2016)\citenamefont {Yokosuk}, \citenamefont {Al-Wahish}, \citenamefont {Artyukhin}, \citenamefont {O'Neal}, \citenamefont {Mazumdar}, \citenamefont {Chen}, \citenamefont {Yang}, \citenamefont {Oh}, \citenamefont {McGill}, \citenamefont {Haule}, \citenamefont {Cheong}, \citenamefont {Vanderbilt},\ and\ \citenamefont {Musfeldt}}]{Yokosuk2016}%
  \BibitemOpen
  \bibfield  {author} {\bibinfo {author} {\bibfnamefont {M.~O.}\ \bibnamefont {Yokosuk}}, \bibinfo {author} {\bibfnamefont {A.}~\bibnamefont {Al-Wahish}}, \bibinfo {author} {\bibfnamefont {S.}~\bibnamefont {Artyukhin}}, \bibinfo {author} {\bibfnamefont {K.~R.}\ \bibnamefont {O'Neal}}, \bibinfo {author} {\bibfnamefont {D.}~\bibnamefont {Mazumdar}}, \bibinfo {author} {\bibfnamefont {P.}~\bibnamefont {Chen}}, \bibinfo {author} {\bibfnamefont {J.}~\bibnamefont {Yang}}, \bibinfo {author} {\bibfnamefont {Y.~S.}\ \bibnamefont {Oh}}, \bibinfo {author} {\bibfnamefont {S.~A.}\ \bibnamefont {McGill}}, \bibinfo {author} {\bibfnamefont {K.}~\bibnamefont {Haule}}, \bibinfo {author} {\bibfnamefont {S.~W.}\ \bibnamefont {Cheong}}, \bibinfo {author} {\bibfnamefont {D.}~\bibnamefont {Vanderbilt}},\ and\ \bibinfo {author} {\bibfnamefont {J.~L.}\ \bibnamefont {Musfeldt}},\ }\bibfield  {title} {\bibinfo {title} {Magnetoelectric coupling through the spin flop transition in {Ni$_3$TeO$_6$}},\ }\href@noop {} {\bibfield  {journal}
  {\bibinfo  {journal} {Physical Review Letters}\ }\textbf {\bibinfo {volume} {117}},\ \bibinfo {pages} {147402} (\bibinfo {year} {2016})}\BibitemShut {NoStop}%
\bibitem [{\citenamefont {Chen}\ \emph {et~al.}(2014)\citenamefont {Chen}, \citenamefont {Holinsworth}, \citenamefont {O'Neal}, \citenamefont {Brinzari}, \citenamefont {Mazumdar}, \citenamefont {Wang}, \citenamefont {McGill}, \citenamefont {Cava}, \citenamefont {Lorenz},\ and\ \citenamefont {Musfeldt}}]{Chen2014}%
  \BibitemOpen
  \bibfield  {author} {\bibinfo {author} {\bibfnamefont {P.}~\bibnamefont {Chen}}, \bibinfo {author} {\bibfnamefont {B.~S.}\ \bibnamefont {Holinsworth}}, \bibinfo {author} {\bibfnamefont {K.~R.}\ \bibnamefont {O'Neal}}, \bibinfo {author} {\bibfnamefont {T.~V.}\ \bibnamefont {Brinzari}}, \bibinfo {author} {\bibfnamefont {D.}~\bibnamefont {Mazumdar}}, \bibinfo {author} {\bibfnamefont {Y.~Q.}\ \bibnamefont {Wang}}, \bibinfo {author} {\bibfnamefont {S.}~\bibnamefont {McGill}}, \bibinfo {author} {\bibfnamefont {R.~J.}\ \bibnamefont {Cava}}, \bibinfo {author} {\bibfnamefont {B.}~\bibnamefont {Lorenz}},\ and\ \bibinfo {author} {\bibfnamefont {J.~L.}\ \bibnamefont {Musfeldt}},\ }\bibfield  {title} {\bibinfo {title} {Magnetic-field-induced shift of the optical band gap in {Ni$_3$V$_2$O$_8$}},\ }\href@noop {} {\bibfield  {journal} {\bibinfo  {journal} {Physical Review B}\ }\textbf {\bibinfo {volume} {89}},\ \bibinfo {pages} {165120} (\bibinfo {year} {2014})}\BibitemShut {NoStop}%
\bibitem [{\citenamefont {Wilson}\ \emph {et~al.}(2021)\citenamefont {Wilson}, \citenamefont {Lee}, \citenamefont {Cenker}, \citenamefont {Xie}, \citenamefont {Dismukes}, \citenamefont {Telford}, \citenamefont {Fonseca}, \citenamefont {Sivakumar}, \citenamefont {Dean}, \citenamefont {Cao}, \citenamefont {Roy}, \citenamefont {Xu},\ and\ \citenamefont {Zhu}}]{Wilson2021}%
  \BibitemOpen
  \bibfield  {author} {\bibinfo {author} {\bibfnamefont {N.~P.}\ \bibnamefont {Wilson}}, \bibinfo {author} {\bibfnamefont {K.}~\bibnamefont {Lee}}, \bibinfo {author} {\bibfnamefont {J.}~\bibnamefont {Cenker}}, \bibinfo {author} {\bibfnamefont {K.}~\bibnamefont {Xie}}, \bibinfo {author} {\bibfnamefont {A.~H.}\ \bibnamefont {Dismukes}}, \bibinfo {author} {\bibfnamefont {E.~J.}\ \bibnamefont {Telford}}, \bibinfo {author} {\bibfnamefont {J.}~\bibnamefont {Fonseca}}, \bibinfo {author} {\bibfnamefont {S.}~\bibnamefont {Sivakumar}}, \bibinfo {author} {\bibfnamefont {C.}~\bibnamefont {Dean}}, \bibinfo {author} {\bibfnamefont {T.}~\bibnamefont {Cao}}, \bibinfo {author} {\bibfnamefont {X.}~\bibnamefont {Roy}}, \bibinfo {author} {\bibfnamefont {X.}~\bibnamefont {Xu}},\ and\ \bibinfo {author} {\bibfnamefont {X.}~\bibnamefont {Zhu}},\ }\bibfield  {title} {\bibinfo {title} {Interlayer electronic coupling on demand in a {2D} magnetic semiconductor},\ }\href@noop {} {\bibfield  {journal} {\bibinfo  {journal} {Nature
  Materials}\ }\textbf {\bibinfo {volume} {20}},\ \bibinfo {pages} {1657} (\bibinfo {year} {2021})}\BibitemShut {NoStop}%
\bibitem [{\citenamefont {Acharya}\ \emph {et~al.}(2023)\citenamefont {Acharya}, \citenamefont {Pashov}, \citenamefont {Weber}, \citenamefont {van Schilfgaarde}, \citenamefont {Lichtenstein},\ and\ \citenamefont {Katsnelson}}]{Acharya2023}%
  \BibitemOpen
  \bibfield  {author} {\bibinfo {author} {\bibfnamefont {S.}~\bibnamefont {Acharya}}, \bibinfo {author} {\bibfnamefont {D.}~\bibnamefont {Pashov}}, \bibinfo {author} {\bibfnamefont {C.}~\bibnamefont {Weber}}, \bibinfo {author} {\bibfnamefont {M.}~\bibnamefont {van Schilfgaarde}}, \bibinfo {author} {\bibfnamefont {A.~I.}\ \bibnamefont {Lichtenstein}},\ and\ \bibinfo {author} {\bibfnamefont {M.~I.}\ \bibnamefont {Katsnelson}},\ }\bibfield  {title} {\bibinfo {title} {A theory for colors of strongly correlated electronic systems},\ }\href@noop {} {\bibfield  {journal} {\bibinfo  {journal} {Nature Communications}\ }\textbf {\bibinfo {volume} {14}},\ \bibinfo {pages} {5565} (\bibinfo {year} {2023})}\BibitemShut {NoStop}%
\bibitem [{\citenamefont {Kanazawa}\ and\ \citenamefont {Street}(1970)}]{Kanazawa1970}%
  \BibitemOpen
  \bibfield  {author} {\bibinfo {author} {\bibfnamefont {K.~K.}\ \bibnamefont {Kanazawa}}\ and\ \bibinfo {author} {\bibfnamefont {G.~B.}\ \bibnamefont {Street}},\ }\bibfield  {title} {\bibinfo {title} {The electrical properties of chromium tribromide},\ }\href@noop {} {\bibfield  {journal} {\bibinfo  {journal} {Physica Status Solidi (b)}\ }\textbf {\bibinfo {volume} {38}},\ \bibinfo {pages} {445} (\bibinfo {year} {1970})}\BibitemShut {NoStop}%
\bibitem [{\citenamefont {Pollini}\ and\ \citenamefont {Spinolo}(1970)}]{Pollini1970}%
  \BibitemOpen
  \bibfield  {author} {\bibinfo {author} {\bibfnamefont {I.}~\bibnamefont {Pollini}}\ and\ \bibinfo {author} {\bibfnamefont {G.}~\bibnamefont {Spinolo}},\ }\bibfield  {title} {\bibinfo {title} {Intrinsic optical properties of {CrCl$_3$}},\ }\href@noop {} {\bibfield  {journal} {\bibinfo  {journal} {Physica Status Solidi (b)}\ }\textbf {\bibinfo {volume} {41}},\ \bibinfo {pages} {691} (\bibinfo {year} {1970})}\BibitemShut {NoStop}%
\bibitem [{\citenamefont {Wang}\ \emph {et~al.}(2018)\citenamefont {Wang}, \citenamefont {Hariki}, \citenamefont {Sotnikov}, \citenamefont {Frati}, \citenamefont {Okamoto}, \citenamefont {Huang}, \citenamefont {Singh}, \citenamefont {Huang}, \citenamefont {Tomiyasu}, \citenamefont {Du}, \citenamefont {Kuneš},\ and\ \citenamefont {Groot}}]{Wang2018}%
  \BibitemOpen
  \bibfield  {author} {\bibinfo {author} {\bibfnamefont {R.~P.}\ \bibnamefont {Wang}}, \bibinfo {author} {\bibfnamefont {A.}~\bibnamefont {Hariki}}, \bibinfo {author} {\bibfnamefont {A.}~\bibnamefont {Sotnikov}}, \bibinfo {author} {\bibfnamefont {F.}~\bibnamefont {Frati}}, \bibinfo {author} {\bibfnamefont {J.}~\bibnamefont {Okamoto}}, \bibinfo {author} {\bibfnamefont {H.~Y.}\ \bibnamefont {Huang}}, \bibinfo {author} {\bibfnamefont {A.}~\bibnamefont {Singh}}, \bibinfo {author} {\bibfnamefont {D.~J.}\ \bibnamefont {Huang}}, \bibinfo {author} {\bibfnamefont {K.}~\bibnamefont {Tomiyasu}}, \bibinfo {author} {\bibfnamefont {C.~H.}\ \bibnamefont {Du}}, \bibinfo {author} {\bibfnamefont {J.}~\bibnamefont {Kuneš}},\ and\ \bibinfo {author} {\bibfnamefont {F.~M.~D.}\ \bibnamefont {Groot}},\ }\bibfield  {title} {\bibinfo {title} {Excitonic dispersion of the intermediate spin state in {LaCoO$_3$} revealed by resonant inelastic x-ray scattering},\ }\href@noop {} {\bibfield  {journal} {\bibinfo  {journal} {Physical Review
  B}\ }\textbf {\bibinfo {volume} {98}} (\bibinfo {year} {2018})}\BibitemShut {NoStop}%
\bibitem [{\citenamefont {Kuindersma}\ \emph {et~al.}(1981)\citenamefont {Kuindersma}, \citenamefont {Boudewijn},\ and\ \citenamefont {Haas}}]{Kuindersma1981a}%
  \BibitemOpen
  \bibfield  {author} {\bibinfo {author} {\bibfnamefont {S.}~\bibnamefont {Kuindersma}}, \bibinfo {author} {\bibfnamefont {P.~R.}\ \bibnamefont {Boudewijn}},\ and\ \bibinfo {author} {\bibfnamefont {C.}~\bibnamefont {Haas}},\ }\bibfield  {title} {\bibinfo {title} {Near-infrared {d-d} transitions of {NiI$_2$}, {CdI$_2$:Ni$^{2+}$}, and {CoI$_2$}},\ }\href@noop {} {\bibfield  {journal} {\bibinfo  {journal} {Phys. Stat. Sol. (b)}\ }\textbf {\bibinfo {volume} {108}},\ \bibinfo {pages} {187} (\bibinfo {year} {1981})}\BibitemShut {NoStop}%
\bibitem [{\citenamefont {He}\ \emph {et~al.}(2025)\citenamefont {He}, \citenamefont {Sears}, \citenamefont {Barantani}, \citenamefont {Kim}, \citenamefont {Villanova}, \citenamefont {Berlijn}, \citenamefont {Lajer}, \citenamefont {McGuire}, \citenamefont {Pelliciari}, \citenamefont {Bisogni}, \citenamefont {Johnston}, \citenamefont {Baldini}, \citenamefont {Mitrano},\ and\ \citenamefont {Dean}}]{He2025}%
  \BibitemOpen
  \bibfield  {author} {\bibinfo {author} {\bibfnamefont {W.}~\bibnamefont {He}}, \bibinfo {author} {\bibfnamefont {J.}~\bibnamefont {Sears}}, \bibinfo {author} {\bibfnamefont {F.}~\bibnamefont {Barantani}}, \bibinfo {author} {\bibfnamefont {T.}~\bibnamefont {Kim}}, \bibinfo {author} {\bibfnamefont {J.~W.}\ \bibnamefont {Villanova}}, \bibinfo {author} {\bibfnamefont {T.}~\bibnamefont {Berlijn}}, \bibinfo {author} {\bibfnamefont {M.}~\bibnamefont {Lajer}}, \bibinfo {author} {\bibfnamefont {M.~A.}\ \bibnamefont {McGuire}}, \bibinfo {author} {\bibfnamefont {J.}~\bibnamefont {Pelliciari}}, \bibinfo {author} {\bibfnamefont {V.}~\bibnamefont {Bisogni}}, \bibinfo {author} {\bibfnamefont {S.}~\bibnamefont {Johnston}}, \bibinfo {author} {\bibfnamefont {E.}~\bibnamefont {Baldini}}, \bibinfo {author} {\bibfnamefont {M.}~\bibnamefont {Mitrano}},\ and\ \bibinfo {author} {\bibfnamefont {M.~P.~M.}\ \bibnamefont {Dean}},\ }\bibfield  {title} {\bibinfo {title} {Dispersive dark excitons in van der waals ferromagnet {CrI$_3$}},\
  }\href@noop {} {\bibfield  {journal} {\bibinfo  {journal} {arXiv}\ ,\ \bibinfo {pages} {2501.09244}} (\bibinfo {year} {2025})}\BibitemShut {NoStop}%
\bibitem [{\citenamefont {Haverkort}(2005)}]{Haverkort2005}%
  \BibitemOpen
  \bibfield  {author} {\bibinfo {author} {\bibfnamefont {M.~W.}\ \bibnamefont {Haverkort}},\ }\bibfield  {title} {\bibinfo {title} {Spin and orbital degrees of freedom in transition metal oxides and oxide thin films studied by soft x-ray absorption spectroscopy},\ }\href@noop {} {\bibfield  {journal} {\bibinfo  {journal} {arXiv}\ ,\ \bibinfo {pages} {0505214}} (\bibinfo {year} {2005})}\BibitemShut {NoStop}%
\end{thebibliography}

%

\end{document}